\documentclass[aps,pra,twocolumn,nopacs,floatfix]{revtex4-2}
\usepackage{epsfig}
\usepackage{graphicx}
\usepackage{dcolumn}
\usepackage{amsthm,amsmath}
\usepackage{comment}
\usepackage[colorlinks]{hyperref}
\usepackage{xcolor}
\usepackage{mathtools}
\usepackage{orcidlink}
\usepackage{amsmath}
\usepackage[normalem]{ulem}

\begin{document}

\title{High-accuracy Nuclear Spin Dependent Parity-Violating Amplitudes in $^{133}$Cs}

\author{$^{a,b}$A. Chakraborty \orcidlink{0000-0001-6255-4584}}
\email{arupc794@gmail.com}

\author{$^a$B. K. Sahoo \orcidlink{0000-0003-4397-7965}}
\email{bijaya@prl.res.in}

\affiliation{
$^a$Atomic, Molecular and Optical Physics Division, Physical Research Laboratory, Navrangpura, Ahmedabad 380009, India}  
\affiliation{
$^b$Indian Institute of Technology Gandhinagar, Palaj, Gandhinagar 382355, India
}

\begin{abstract}
Relativistic coupled-cluster (RCC) theory at the singles and doubles approximation has been implemented to estimate nuclear spin dependent (NSD) parity-violating (PV) electric dipole (E1) transition amplitudes ($E1_{PV}^{NSD}$) among hyperfine levels of the $6s ~^2S_{1/2} \rightarrow 7s ~^2S_{1/2}$ transition in $^{133}$Cs. To validate our calculations, we reproduce the Dirac-Hartree-Fock values and results from the combined coupled-Dirac-Hartree-Fock and random phase approximation (CPDF-RPA) method reported earlier. Contributions from the double-core-polarization (DCP) effects at the CPDF-RPA method were found to be between 3-12\% among different hyperfine levels. We derived a generalized expression for $E1_{PV}^{NSD}$, which helped incorporate both the NSD PV Hamiltonian and E1 operator simultaneously in the perturbative approach to account for the DCP contributions. The RCC method subsumes the CPDF-RPA and DCP effects in addition to contributions from the Br\"uckner pair-correlations and normalization of the wave functions, and correlations among them. To improve accuracy of the $E1_{PV}^{NSD}$ amplitudes further, we replace the {\it ab initio} values of the E1 matrix elements and energies by their experimental values via a sum-over-states approach.
\end{abstract}

\date{\today}

\maketitle 

\section{Introduction}
In an atomic system, parity violation (PV) interactions originate from two primary sources \cite{Bouchiat1997, Commins1983}. The first source is the neutral current weak interactions between the atomic nucleus and electrons. The second is the electromagnetic interaction between electrons and a potentially existing parity-violating nuclear anapole moment (NAM) within the nucleus. Though the concept of NAM is very fundamental, its existence is still a debate and should be addressed decisively. The weak interactions, mediated by the $Z_0$ boson (and $Z_0$-like bosons), are categorized into nuclear spin-independent (NSI) and nuclear spin-dependent (NSD) interactions, depending on whether the axial-vector and vector currents arise from the electron and nuclear sectors, respectively, or vice versa \cite{Bouchiat1997}. The NSI component, owing to the coherent contributions from all the nucleons, is sensitive to probe possible new physics beyond the Standard Model (SM) of particle interactions \cite{Ginges2004}. The PV interactions in atomic systems are too weak to be detected directly like other typical spectroscopic properties. In view of this, special techniques have been developed in different laboratories to observe these effects. Usually, electric dipole (E1) amplitudes of the forbidden transitions due to PV effects ($E1_{PV}$) in combinations with either magnetic dipole amplitudes, electric quadrupole amplitudes or Stark-induced E1 amplitudes ($E1_{Stark}$) are measured to realise their signatures in atomic systems. Among a few selective atomic systems, $^{133}$Cs was considered by two groups to measure $E1_{PV}$ \cite{Bouchiat1975, Lintz2007, Wood1997}.  Both the groups considered the $6s ~ ^2S_{1/2} \rightarrow 7s ~ ^2S_{1/2}$ transition in $^{133}$Cs in their experiments, but one group measured the interface between the $E1_{PV}$ and optical transition amplitudes  \cite{Lintz2007} while the other group has reported measurement of the interface between the $E1_{PV}$ and $E1_{Stark}$ amplitudes \cite{Wood1997}. To differentiate the NSI and NSD components from $E1_{PV}$, denoted by $E1_{PV}^{NSI}$ and $E1_{PV}^{NSD}$ respectively, from the experimental results, measurements were carried out between the $F=3$ and $F=4$ hyperfine levels of the $6s ~ ^2S_{1/2}$  and $7s ~ ^2S_{1/2}$ states by Wood et al. \cite{Wood1997}. This offers experimental values to the NSI and NSD components of $E1_{PV}$ in $^{133}$Cs within 0.35\% and 15\% accuracy, respectively, \cite{Wood1997}. This is the most accurate measurement of $E1_{PV}$ in an atomic system till date.

Several groups have put continuous efforts into determining the $E1_{PV}^{NSI}$ amplitude of the $6s ~ ^2S_{1/2} \rightarrow 7s ~ ^2S_{1/2}$ transition in $^{133}$Cs for the last two decades. Among these, calculations performed using the relativistic coupled-cluster (RCC) theory are considered to be more accurate \cite{Porsev2010, Dzuba2012, Sahoo2021}. Combining the calculations with the precise measurement of $E1_{PV}^{NSI}$, stringent constraints on several fundamental parameters signifying beyond the SM physics have been imposed. However, less effort has been put into calculating the $E1_{PV}^{NSD}$ amplitude of the above transition. This is owing to the fact that the electronic component of the NSD PV interaction Hamiltonian is a rank one operator and its coupling with nuclear spin makes angular momentum couplings complicated compared to calculations involving the NSI interaction Hamiltonian. Though a finite value of NAM has been inferred by combining the measured $E1_{PV}^{NSD}$ amplitude with the earlier calculations, the inferred value is at variance with the results of the shell model and the nucleon-nucleon scattering experiments \cite{Wilburn1998, Haxton2001}. Moreover, the sign of the NAM coupling constant from the Cs measurement is not in agreement with the measurement using the Tl atom \cite{Ginges2004}. In such circumstances, it is imperative to carry out further investigations on NAM. This motivated us to revisit the calculation of $E1_{PV}^{NSD}$ in the $6s ~ ^2S_{1/2} \rightarrow 7s ~ ^2S_{1/2}$ transition of $^{133}$Cs by employing the RCC theory. 

In one of the early calculations, Flambaum and Dzuba had used the Brueckner orbitals (correspond to pair-correlation (PC) effects) in the Dirac-Hartree-Fock (DHF) method to estimate the $E1_{PV}^{NSD}$ values between different hyperfine levels of the $6s ~ ^2S_{1/2} \rightarrow 7s ~ ^2S_{1/2}$ transition of $^{133}$Cs \cite{Dzuba1987, Flambaum1997}. Subsequently, Johnson et al. employed random phase approximation (RPA) to calculate these amplitudes by incorporating electron correlation effects due to core-polarization (CP) effects to all-order \cite{Johnson2003}. As demonstrated in the calculation of $E1_{PV}^{NSI}$, the non-RPA correlation effects are quite significant in the $6s ~ ^2S_{1/2} \rightarrow 7s ~ ^2S_{1/2}$ transition of $^{133}$Cs \cite{Chakraborty2023}. Moreover, the double core-polarization (DCP) contributions, the CP effects arising due to simultaneous consideration of both PV interaction Hamiltonian and E1 operator in the perturbation, to $E1_{PV}^{NSD}$ in the above transition are not yet explored explicitly like it is investigated for $E1_{PV}^{NSI}$ \cite{Chakraborty2023, Roberts2013}. In order to address disagreement between the nuclear theory probe and atomic studies on the NAM, it is necessary to calculate the $E1_{PV}^{NSD}$ amplitudes in $^{133}$Cs more accurately by including the correlation effects that were omitted earlier. In this work, we estimate the $E1_{PV}^{NSD}$ amplitudes between different hyperfine levels of the $6s ~ ^2S_{1/2}$ and $7s ~ ^2S_{1/2}$ states by employing the RCC theory, which captures the PC, CP and DCP contributions to all-orders. To validate our calculations and show the importance of considering the DCP effects for accurate determination of the $E1_{PV}^{NSD}$ amplitudes, we first analyze the results using the DHF, RPA, coupled-Dirac-Hartree-Fock (CPDF) and combined CPDF-RPA methods. These results are compared with the previously reported values at the same level of approximation in the many-body theory. Since the commonly used formula to estimate the $E1_{PV}^{NSD}$ amplitude in atomic system is not apt to use for estimating the DCP effect in the CPDF-RPA approach, we present a generic expression that can be implemented easily in any of the many-body method. In addition, we improve the RCC results by replacing {\it ab initio} values of the E1 amplitudes and energies by their experimental results via a sum-over-states approach.

\section{Theory} \label{secth}

The NSD PV interaction Hamiltonian in an atomic system is given by 
\begin{eqnarray}
H_{NSD} &=& K_W \frac{G_F}{\sqrt 2 } \left ( {\vec \alpha}^D \cdot {\vec I} \right ) \rho_N(r) ,
\end{eqnarray}
where $G_F$ is the Fermi constant with value $2.22249 \times 10^{-14}$ in atomic units (a.u.), the dimensionless quantity $K_W$ is a NSD PV violating nuclear parameter, ${\vec \alpha}^D$ is the Dirac matrix, ${\vec I}$ is the nuclear spin and $\rho_N(r)$ is the nuclear density. The magnitude of $K_W$ depends on the contributions from NAM and NSD interactions within an atomic nucleus, so it can be expressed as
\begin{eqnarray}
K_W=K_a+K_{NSD} ,    
\end{eqnarray}
where $K_a$ and $K_{NSD}$ denote contributions from NAM and NSD interactions, respectively. To determine the nuclear potential and density, we have used Fermi charge distribution, given by
\begin{equation}
\rho_N(r)=\frac{\rho_{0}}{1+e^{(r-b)/a}} 
\end{equation}
with the normalization factor $\rho_0$, the half-charge radius $b=5.670729105$ fm \cite{Angeli2004} and $a= 2.3/4({\rm ln}3)$ is related to the skin thickness. As $K_W$ is the quantity to be inferred by combining measurement with atomic calculation, we express the above Hamiltonian  as
\begin{eqnarray}
H_{NSD} &=& K_W  H_{PV}^{NSD} ,
\end{eqnarray}
so that calculations can be performed using $H_{PV}^{NSD}$. Again, it would be necessary to decouple the electronic and nuclear components from $H_{PV}^{NSD}$ in order to facilitate the calculations using the electronic part. This is done by expressing 
\begin{eqnarray}
 H_{PV}^{NSD} &=& \frac{G_F}{\sqrt{2}} \sum_q (-1)^q I_q^{(1)} K_{-q}^{(1)} ,
\label{eq2}
\end{eqnarray}
where $I_q^{(1)}$ is the $q$th component of $\vec{I}$ and the matrix element of the electronic component, $K^{(1)}$, between the orbitals $|\phi_f \rangle$ and $|\phi_i\rangle$ in the relativistic form is given by
\begin{eqnarray}
\langle \phi_f | K_q^{(1)} | \phi_i \rangle &=&  i \int_0^{\infty} dr \rho_N(r) \nonumber \\
 && \times \left [ \langle \kappa_f m_f | \sigma_q | - \kappa_i m_i \rangle P_f(r) Q_i (r) \right. \nonumber \\
&&  \left. - \langle - \kappa_f m_f | \sigma_q | \kappa_i m_i \rangle Q_f(r) P_i (r) \right ].
\end{eqnarray} 
In the above expression, $P(r)$ and $Q(r)$ represent the large and small components of the Dirac wave function respectively, $\kappa$ and $m$ are the relativistic and azimuthal component of the angular momentum of the Dirac orbital respectively, and $\sigma$ is the Pauli spinor.

In the presence of $H_{NSD}$, a mixed-parity hyperfine level wave function, $|\Psi_v \rangle^{F}$, can be expressed as 
\begin{eqnarray}
|\Psi_v \rangle^{F} \simeq |\Psi_v^{F,(0)} \rangle + K_W | \Psi_v^{F,(1)} \rangle ,
\end{eqnarray}
where $|\Psi_v^{F,(0)} \rangle$ is the hyperfine level wave function due to electromagnetic interactions and $|\Psi_v^{F,(1)} \rangle$ is the first correction due to $H_{PV}^{NSD}$. Note that $K_W$ may not be small, but $H_{PV}^{NSD}$ is very small compared to the Hamiltonian describing electromagnetic interactions in the atomic systems. So in the above perturbative analysis $K_W$ just denotes the order of perturbation rather than the strength of the interaction.

Between the hyperfine levels with $|\Psi_f\rangle^{F} \equiv |(I J_f) F_f M_f \rangle$ and $|\Psi_i \rangle^{F} \equiv |(I J_i) F_i M_i \rangle$, the $E1_{PV}^{NSD}$ can be given using the Wigner-Eckart theorem as
\begin{eqnarray}
E1_{PV}^{NSD}  &=& (-1)^{F_f-M_f}\begin{pmatrix}
  F_f  & 1 & F_i\\
  -M_f & M_f - M_i &  M_i
 \end{pmatrix} \nonumber \\ && \times \langle F_f || D_{PV}^{NSD} || F_i \rangle , 
 \label{eqhf}
\end{eqnarray}
where $\langle F_f || D_{PV}^{NSD} || F_i \rangle$ is the reduced matrix element and $D_{PV}^{NSD}$ is the PV interaction induced E1 operator. The actual quantity of interest from atomic calculation point of view is ${\cal X}_{PV}^{NSD} = \langle F_f || D_{PV}^{NSD} || F_i \rangle / K_W$. 

By expanding the first-order wave function in the sum-over-states approach, we can write
\begin{eqnarray}
{\cal X}_{PV}^{NSD} &=&  \sum_{n \ne i} \frac{ \langle F_f || D || F_n \rangle \langle F_n || H_{PV}^{NSD} || F_i \rangle} { {\cal N}_F (E_{F_i}^{(0)} - E_{F_n}^{(0)}) } \nonumber \\  &+&
\sum_{n \ne f} \frac{\langle F_f || H_{PV}^{NSD} || F_n \rangle \langle F_n || D|| F_i \rangle} { {\cal N}_F (E_{F_f}^{(0)} - E_{F_n}^{(0)}) }  \nonumber \\
 & \simeq &  \sum_{n \ne i} \frac{ \langle F_f || D || F_n \rangle \langle F_n || H_{PV}^{NSD} || F_i \rangle} { {\cal N}_F (E_i^{(0)} - E_n^{(0)}) } \nonumber \\  &+&
\sum_{n \ne f} \frac{\langle F_f || H_{PV}^{NSD} || F_n \rangle \langle F_n || D|| F_i \rangle} { {\cal N}_F (E_f^{(0)} - E_n^{(0)}) },
\label{eq3}
\end{eqnarray}
where $D$ is the usual E1 operator, $E_{F_n}^{(0)}$ and $E_n^{(0)}$ are the hyperfine and atomic energy values of the $n^{th}$ state, respectively, and ${\cal N}_F = \sqrt{\langle \Psi_f^{(0)} | \Psi_f^{(0)} \rangle^F \langle \Psi_i^{(0)} | \Psi_i^{(0)} \rangle^F }$ is the normalization factor of the hyperfine level states. It is challenging to deal with the wave functions in the hyperfine coordinate system to evaluate the above quantity. To address this, we express the $|(IJ)F M_F\rangle$ levels in perturbation series as 
\begin{eqnarray}
|(IJ)F M_F \rangle &=& |II ; J M_J \rangle + \sum_{J',M_{J'}} |II ; J' M_{J'} \rangle\nonumber\\
&&\times \frac{\langle II ; J' M_{J'} | H_{hf} |I I ; J M_J \rangle} {E_J - E_{J'}} + \cdots.
\label{eqrephf}
\end{eqnarray}
In the above expression, $H_{hf}$ denotes the hyperfine interaction Hamiltonian. In this work, we restricted ourselves to only the first term $|II; J M_J \rangle$. Using the above approximation and substituting the following relations 
\begin{eqnarray}\label{eqme1}
\langle (I,J_n) F_n, M_n | \vec{K}^{(1)}\cdot \vec{I}| (I,J_i) F_i,M_i \rangle = \delta_{F_n,F_i}\delta_{M_n,M_i} \nonumber \\
 \times (-1)^{I+F_i+J_i}\sqrt{I(I+1)(2I+1)} \nonumber\\
\times  \begin{Bmatrix} 
     J_n & J_i & 1\\
     I   &  I  & F_i
   \end{Bmatrix}\langle J_n||K^{(1)}||J_i\rangle \ \ \ \
\end{eqnarray}
 and 
\begin{eqnarray}
\langle (I,J_f) F_f,M_f| D| (I,J_n) F_n,M_n \rangle = \sqrt{(2F_f+1)(2F_n+1)}
\nonumber\\
 \times (-1)^{F_f-M_f} \begin{pmatrix}
  F_f  & 1 & F_n\\
  -M_f & q &  M_n
\end{pmatrix} (-1)^{I+F_n+J_f+1} \nonumber \\
\times    \begin{Bmatrix} 
     J_n & J_f & 1\\
     F_f   & F_n  & I
   \end{Bmatrix} \langle J_f||D||J_n\rangle \ \ \ \ \ \ \ 
\end{eqnarray}
in Eq. (\ref{eq3}), it yields
\begin{eqnarray}\label{eqNSD1}
{\cal X}_{PV}^{NSD}&=& {\cal C} \bigg[ \sum_{n\ne i}  \begin{Bmatrix} 
     J_n & J_i & 1\\
     I   &  I  & F_i
   \end{Bmatrix}  \begin{Bmatrix} 
     J_n & J_f & 1\\
     F_f   & F_i  & I
   \end{Bmatrix} \nonumber \\ \times && (-1)^{(J_f-J_i+1)} 
   \frac{\langle J_f||D||J_n\rangle\langle J_n||K^{(1)}||J_i\rangle }{E^{(0)}_i-E^{(0)}_n}\nonumber\\
   &+& \sum_{n\ne f}  \begin{Bmatrix} 
     J_n & J_f & 1\\
     I   &  I  & F_f
   \end{Bmatrix}  \begin{Bmatrix} 
     J_n & J_i & 1\\
     F_i   & F_f  & I
   \end{Bmatrix} \nonumber \\ \times && (-1)^{(F_f-F_i+1)} 
   \frac{\langle J_f||K^{(1)}||J_n\rangle\langle J_n||D||J_i\rangle }{E^{(0)}_f-E^{(0)}_n}\bigg] , \ \ \
   \label{eq4}
\end{eqnarray} 
where ${\cal C}= \frac{G_F}{\sqrt2}\sqrt{I(I+1)(2I+1)(2F_f+1)(2F_i+1)}$ .

To carry out calculation of the above quantity using the final state ($|\Psi_f\rangle$) and initial state ($|\Psi_i \rangle$) atomic wave functions, we can express Eq. (\ref{eq4}) as 
\begin{eqnarray}
{\cal X}_{PV}^{NSD} & = &  \frac{1}{{\cal N}} \langle \Psi_f^{(0)} | \tilde{D}_f | \Psi_i^{(1)} \rangle  + \langle \Psi_f^{(1)} | \tilde{D}_i | \Psi_i^{(0)} \rangle , 
\end{eqnarray}
where ${\cal N} = \sqrt{\langle \Psi_f^{(0)} | \Psi_f^{(0)} \rangle \langle \Psi_i^{(0)} | \Psi_i^{(0)} \rangle }$ is the normalization factor of the atomic wave functions, $\tilde{D}_i$ and $\tilde{D}_f$ are the effective E1 operators, the wave functions with superscript $(0)$ and ($1$) are the unperturbed atomic wave functions and first-order perturbed atomic wave functions due to the $K^{(1)}$ operator corresponding to the initial and final states respectively. The effective E1 operators are explicitly given by 
\begin{eqnarray}
\tilde{D}_i &=&  (-1)^{(F_f-F_i+1)} \sum_n \langle J_n||D||J_i\rangle \nonumber \\
\times && \begin{Bmatrix} 
     J_n & J_f & 1\\
     I   &  I  & F_f
   \end{Bmatrix}  \begin{Bmatrix} 
     J_n & J_i & 1\\
     F_i   & F_f  & I
   \end{Bmatrix} 
\end{eqnarray}
and
\begin{eqnarray}
\tilde{D}_f &=& (-1)^{(J_f-J_i+1)} \sum_n  \langle J_f||D||J_n\rangle \nonumber \\
\times && \begin{Bmatrix} 
     J_n & J_i & 1\\
     I   &  I  & F_i
   \end{Bmatrix}  \begin{Bmatrix} 
     J_n & J_f & 1\\
     F_f   & F_i  & I
   \end{Bmatrix} .
\end{eqnarray}
The first-order atomic wave functions are obtained by solving the following inhomogeneous equations
\begin{eqnarray}
 (H_{at} - E_v^{(0)} ) |\Psi_v^{(1)} \rangle = -K^{(1)} |\Psi_v^{(0)} \rangle ,
\end{eqnarray}
where $H_{at}$ denotes atomic Hamiltonian due to electromagnetic interactions.

Alternatively, the ${\cal X}_{PV}^{NSD}$ amplitude between the states $|\Psi_f\rangle$ and $|\Psi_i \rangle$ can be estimated as the second-order correction after considering E1 operator, $D$, as an additional perturbation (similar to the expressions given for $E1_{PV}^{NSI}$ in Ref. \cite{Chakraborty2023}). i.e.
\begin{eqnarray}
{\cal X}_{PV}^{NSD} & \simeq & \langle \Psi_f^{(0,0)} | \overline{D} | \Psi_i^{(1,0)} \rangle  + \langle  \Psi_f^{(0,0)} | \overline{K}^{(1)} |  \Psi_i^{(0,1)} \rangle \nonumber \\ 
&&  + \langle \Psi_f^{(0,0)} | \Psi_i^{(1,1)} \rangle ,
\label{eq5}
\end{eqnarray}
in which the superscripts $(m,n)$ denotes $m$ orders of $K^{(1)}$ and $n$ orders of $D$. For example, Eq. (\ref{eq3}) is used in the RPA, CPDF and CPDF-RPA methods without considering the DCP effects. However, the DCP effects can be considered in the CPDF-RPA method through the last term of Eq. (\ref{eq5}). The RCC theory can include the DCP effects through the formulation of the method either adopting Eq. (\ref{eq3}) or  Eq. (\ref{eq5}). It is not convenient to determine the second-order perturbed wave function, $| \Psi_i^{(1,1)} \rangle$, using the formula given by Eq. (\ref{eq4}). For this, we give here a more generic formula by simplifying $\langle F_f || D_{PV}^{NSD} || F_i \rangle$ using the tensor product relations \cite{Lindgren1985, Varshalovich1989} as
\begin{eqnarray}
{\cal X}_{PV}^{NSD} &=& {\cal C} \sum_{k=0,1,2} (2k+1) \nonumber \\ && \times \bigg[ \sum_{n\ne i}  \begin{Bmatrix} 
     J_f & J_i & k\\
     I   &  I  & 1\\
     F_f & F_i & 1
   \end{Bmatrix}  \begin{Bmatrix} 
     J_f & k & J_i\\
     1   & J_n & 1
   \end{Bmatrix}\nonumber\\
   &\times&(-1)^{(J_f+J_i+1)}\frac{\langle J_f||D||J_n\rangle\langle J_n||K^{(1)}||J_i\rangle }{E^{(0)}_i-E^{(0)}_n}\nonumber\\
   &+& \sum_{n\ne f} \begin{Bmatrix} 
    I   &  I  & 1\\
    J_f & J_i & k\\
     F_f & F_i & 1
   \end{Bmatrix}   \begin{Bmatrix} 
     J_f & k & J_i\\
     1   & J_n & 1
   \end{Bmatrix}\nonumber\\
      &\times&(-1)^{(2J_f+F_f-F_i+1)} \frac{\langle J_f||K^{(1)}||J_n\rangle\langle J_n||D||J_i\rangle }{E^{(0)}_f-E^{(0)}_n}\bigg] . \ \ \ \ \
   \label{eq6} 
\end{eqnarray}
It can be shown that both Eqs. (\ref{eq4}) and (\ref{eq6}) are equivalent. However, $F_i$, $F_f$ and $J_n$ are not coupled through either $6j$- or $9j$-symbols in Eq. (\ref{eq6}) as in the case for Eq. (\ref{eq4}). This helps to implement the above expression in the CPDF-RPA method to compute ${\cal X}_{PV}^{NSD}$ by defining $ \overline{D} $ and $\overline{K}^{(1)}$ as
\begin{eqnarray}
\overline{D} &=& (-1)^{(J_f+J_i+1)} \sum_{k=0,1,2} (2k+1)   \nonumber \\ \times && \sum_{n}  \begin{Bmatrix} 
     J_f & J_i & k\\
     I   &  I  & 1\\
     F_f & F_i & 1
   \end{Bmatrix}  \begin{Bmatrix} 
     J_f & k & J_i\\
     1   & J_n & 1
   \end{Bmatrix} \langle J_f||D||J_n\rangle \ \ \ \
\end{eqnarray}
and
\begin{eqnarray}
\overline{K}^{(1)} &=& (-1)^{(2J_f+F_f-F_i+1)} \sum_{k=0,1,2} (2k+1) \nonumber \\ \times &&  \sum_{n}  
\begin{Bmatrix} 
    I   &  I  & 1\\
    J_f & J_i & k\\
     F_f & F_i & 1
   \end{Bmatrix}   \begin{Bmatrix} 
     J_f & k & J_i\\
     1   & J_n & 1
   \end{Bmatrix} \langle J_f||K^{(1)}||J_n\rangle . \ \ \ \
\end{eqnarray}
In this case, the first-order wave functions are evaluated in the first-principle approach as 
\begin{eqnarray}
 (H_{at} - E_i^{(0)} ) |\Psi_i^{(1,0)} \rangle = -K^{(1)} |\Psi_i^{(0,0)} \rangle 
\end{eqnarray}
 and
 \begin{eqnarray}
 (H_{at} - E_i^{(0)}-\omega)| \Psi_i^{(0,1)} \rangle = -D |\Psi_i^{(0,0)} \rangle,
\end{eqnarray}
where $\omega=E_f^{(0)}-E_i^{(0)}$. It should be noted that the matrix elements of both the $K^{(1)}$ and $D$ operators and all the coupling angular factors are taken into account in the evaluation of $| \Psi_i^{(1,1)} \rangle$, so it does not require to define any additional effective operator for estimating ${\cal X}_{PV}^{NSD}$.

\section{Methodology}

\begin{table}[t]
\caption{Estimated ${\cal X}_{PV}^{NSD}$ values of the hyperfine transitions ($F_f - F_i$) among all possible hyperfine levels $F_f$ and $F_i$ of the $7s ~ ^2S_{1/2}$ and $6s ~ ^2S_{1/2}$ states, respectively, in $^{133}$Cs from different methods. All the values given in the units of $i e a_0 K_W\times10^{-12}$ with the electron charge $e$ and the Bohr radius $a_0$. We also compare these values with the values reported in previous works at different levels of approximation in the many-body method.}
\begin{ruledtabular}
\begin{tabular}{lcccc}
  Method  &  $3-3$ & $3-4$ & $4 - 3$   & $4 - 4$ \\
\hline \\[1ex]
\multicolumn{5}{c}{This work}\\[0.3ex]
DHF & 1.9029 &	5.4663 & 4.7337 & 2.1665\\
CPDF & 2.3345 &	7.0455 & 6.1470 & 2.6579\\
RPA & 1.8305 & 5.6738 & 4.9689 & 2.0842\\
CPDF-RPA* & 2.2456 & 7.2348 & 6.3707 & 2.5564\\
CPDF-RPA & 2.0139 & 6.9891 & 6.2142 & 2.2928\\
RCCSD & 2.3344 & 7.3943 & 6.4958 & 2.6575\\[2ex]
\multicolumn{5}{c}{Other works}\\[0.5ex]

DHF \cite{Johnson2003} & 1.908 & 5.481 &  4.746 & 2.173 \\ [0.5ex]
DHF \cite{Mani2011}  & 2.011 & 5.774 & 5.000 & 2.289 \\ [0.5ex]

RPA \cite{Johnson2003} &  2.249 & 7.299 & 6.432 & 2.560 \\[0.5ex]

PRCC$^{\dagger}$ \cite{Mani2011}  & 2.274 & 6.313 & 5.446 & 2.589 \\[0.5ex]
 
SD \cite{Safronova2009} &  & 7.948 & 7.057 &  \\[0.5ex]
\end{tabular}
\end{ruledtabular} 
\begin{flushleft}
$^{\dagger}$The PRCC method of Ref. \cite{Mani2011} is same as our RCCSD method.    
\end{flushleft}
\label{tab0}
\end{table} 

To determine the unperturbed electronic wave functions of the atomic states of $^{133}$Cs, we consider $H_{at}$ at the Dirac-Coulomb (DC) approximation, given in atomic units (a.u.) by
\begin{eqnarray}\label{eq:DC}
H_{at} &=& \sum_i \left [c {\vec \alpha}_i^D \cdot {\vec p}_i+(\beta_i^D-1)c^2+V_n(r_i)\right ] \nonumber \\
&& +\sum_{i,j>i}\frac{1}{r_{ij}}, 
\end{eqnarray}
where $c$ is speed of light, $\beta^D$ is another Dirac matrix, ${\vec p}$ is the single particle momentum operator, $V_n(r)$ denotes nuclear potential seen by an electron at distance $r$ from the nucleus and $\frac{1}{r_{ij}}$ represents the Coulomb potential between the electrons located at the $i^{th}$ and $j^{th}$ positions. 

The final unperturbed wave functions are obtained in three steps. In the first step we determine the Dirac-Hartree-Fock (DHF) wave function, $|\Phi_0 \rangle$, due to  $H_{at}$ of the closed-shell core $[5p^6]$ of $^{133}$Cs. In the second step, exact atomic wave function of the closed-core, $|\Psi_0^{(0)} \rangle$ is determined by incorporating the electron correlation effects due to the residual Coulomb interactions, $V_{es}=H_{at}-H_{DHF}$ with the DHF Hamiltonian $H_{DHF}$, neglected in the DHF method using an wave operator $\Omega_0^{(0)}$, i.e.
\begin{eqnarray}
|\Psi_0^{(0)} \rangle = \Omega_0^{(0)} |\Phi_0 \rangle .
\end{eqnarray}

\begin{figure}[t!]
\centering
\includegraphics[height=80mm,width=80mm]{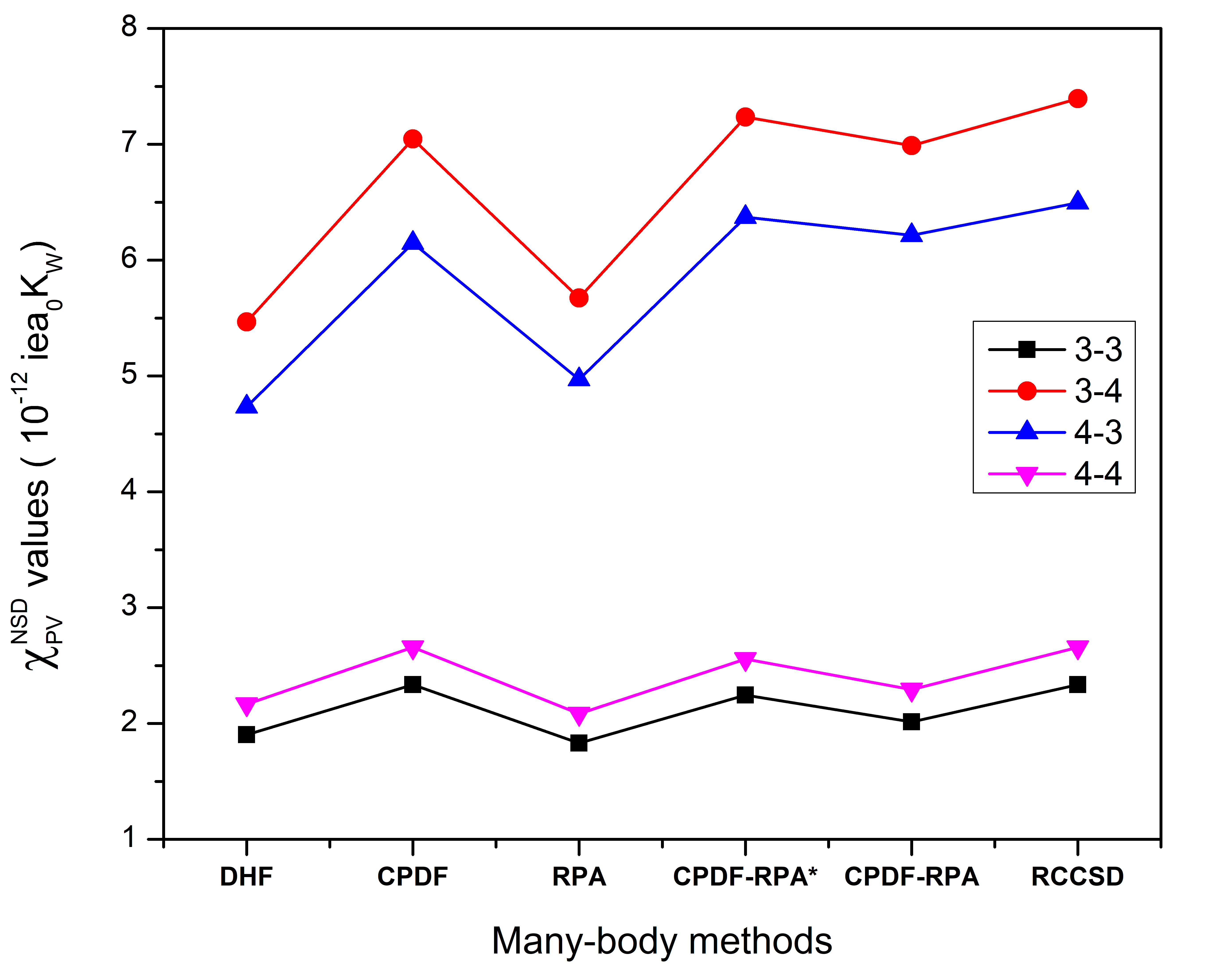}
\caption{Magnitudes of the ${\cal X}_{PV}^{NSD}$ values for the $F_f-F_i$ transitions of the $7s ~ ^2S_{1/2}(F_f)$ and $6s ~ ^2S_{1/2}(F_i)$ states in $^{133}$Cs from different methods. Same units as in Table \ref{tab0} are used for the plotted values.}
\label{fig0}
\end{figure}
 
In the third and final step, we obtain the intended wave functions of $^{133}$Cs by appending the required valence orbital, $v$, of the state to the closed-core configuration $[5p^6]$. For this purpose, the modified DHF wave function is defined as $|\Phi_v \rangle = a_v^{\dagger} |\Phi_0 \rangle$. The exact unperturbed wave function of state, $|\Psi_v^{(0)} \rangle$, in such case can be defined as
\begin{eqnarray}
|\Psi_v^{(0)} \rangle = (\Omega_0^{(0)} + \Omega_v^{(0)}) |\Phi_v \rangle,
\end{eqnarray}
where $\Omega_v^{(0)}$ is the new wave operator responsible for exciting electrons including the valence electron from $|\Phi_v \rangle$. It should be noted that $\Omega_0^{(0)}$ excites electrons only from the core orbitals of  $|\Phi_v \rangle$. It means that we need to determine amplitudes of both the operators to determine $|\Psi_v^{(0)} \rangle$. 

Analogous to unperturbed wave functions, we can also define wave operators to obtain the first-order perturbed wave functions due to $H_{NSD}$ as
\begin{eqnarray}
|\Psi_0^{(1)} \rangle = \Omega_0^{(1)} |\Phi_0 \rangle 
\end{eqnarray}
and
\begin{eqnarray}
|\Psi_v^{(1)} \rangle = (\Omega_0^{(1)} +  \Omega_v^{(1)}) |\Phi_v \rangle ,
\end{eqnarray}
where superscript $(1)$ denotes for the first-order perturbation. When both the $H_{NSD}$ and $D$ operators are included one-order each, then the second-order perturbed wave functions are defined as
\begin{eqnarray}
|\Psi_0^{(1,1)} \rangle = \Omega_0^{(1,1)} |\Phi_0 \rangle 
\end{eqnarray}
and
\begin{eqnarray}
|\Psi_v^{(1,1)} \rangle = (\Omega_0^{(1,1)} +  \Omega_v^{(1,1)} ) |\Phi_v \rangle ,
\end{eqnarray}
where the first superscript stands for the order of $H_{NSD}$ and the second superscript stands for the order of $D$.

The DHF expression for $E1_{PV}^{NSD}$ using Eq. (\ref{eq3}) can be given by \cite{Chakraborty2023}
\begin{eqnarray}
{\cal X}_{PV}^{NSD}  &=& {\cal C} \langle \Phi_f | \Omega_0^{DHF,(0) \dagger} \tilde{D}_f \Omega_0^{DHF,(1) }| \Phi_i \rangle \nonumber \\
&& + \langle \Phi_f | \Omega_0^{DHF,(1) \dagger}\tilde{D}_i \Omega_0^{DHF,(0) } | \Phi_i \rangle \nonumber \\
&& + \langle \Phi_f | \Omega_f^{DHF,(0) \dagger}\tilde{D}_f \Omega_i^{DHF,(1) } | \Phi_i \rangle \nonumber \\
&& + \langle \Phi_f | \Omega_f^{DHF,(1) \dagger} \tilde{D}_i \Omega_i^{DHF,(0) } | \Phi_i \rangle,
\label{eqnhf}
\end{eqnarray}
where $\Omega_0^{DHF,(0) } = \Omega_v^{DHF,(0) } =1$, $\Omega_0^{DHF,(1)} = \sum_{ap} \frac{ \langle \Phi_a^p | K^{(1)} | \Phi_0 \rangle} { {\cal E}_0 + \epsilon_p -{\cal E}_0 - \epsilon_a} a_p^{\dagger} a_a$ and $\Omega_v^{DHF,(1)} = \sum_p \frac{ \langle \Phi_v^p | K^{(1)} | \Phi_v \rangle} { {\cal E}_0 + \epsilon_p - {\cal E}_0 - \epsilon_v} a_p^{\dagger} a_v$ with ${\cal E}_0=\sum_a \epsilon_a$ is the DHF energy of the closed-core and $\epsilon_i$ is the $i^{th}$ DHF orbital. Here, $a,b$ denote for core orbitals, $p,q$ denote for virtual orbitals and $|\Phi_{ab\cdots}^{pq\cdots} \rangle = a_p^{\dagger} a_q^{\dagger} \cdots a_b a_a |\Phi_0 \rangle$. We have also added name of the method in the superscript of the wave operator in order to identify the approximation used in the calculation. It should be noted that in Eq. (\ref{eqnhf}), contributions from the first two terms are referred to as the ``Core" contribution while the contributions from the last two terms are called as ``Valence" contributions. 

In the CPDF method, Eq. (\ref{eq3}) is again used to calculate ${\cal X}_{PV}^{NSD}$. We express its formula as \cite{Chakraborty2023}
\begin{eqnarray}
{\cal X}_{PV}^{NSD} &=& {\cal C} \langle \Phi_f | \Omega_0^{CPDF,(0) \dagger} \tilde{D}_f \Omega_0^{CPDF,(1) } | \Phi_i \rangle \nonumber \\
&& + \langle \Phi_f | \Omega_0^{CPDF,(1) \dagger} \tilde{D}_i \Omega_0^{CPDF,(0) } | \Phi_i \rangle \nonumber \\
&& + \langle \Phi_f | \Omega_f^{CPDF,(0) \dagger} \tilde{D}_f \Omega_i^{CPDF,(1) } | \Phi_i \rangle \nonumber \\
&& + \langle \Phi_f | \Omega_f^{CPDF,(1) \dagger} \tilde{D}_i \Omega_i^{CPDF,(0) } | \Phi_i \rangle,
\label{eqncpdf1}
\end{eqnarray}
where $\Omega_{0/v}^{CPDF,(0)}\equiv\Omega_{0/v}^{DHF,(0)}$ and the amplitudes of the first-order CPDF wave operators are obtained by 
\begin{eqnarray}
(H_{DHF} - {\cal E}_0) \Omega_0^{CPDF,(1)} |\Phi_0 \rangle =  - K^{(1)} | \Phi_0 \rangle - U_{PV}^{(1)}  |\Phi_0 \rangle  \nonumber 
\end{eqnarray}
and
\begin{eqnarray}
(H_{DHF} - {\cal E}_v) \Omega_v^{CPDF,(1) } |\Phi_v \rangle =  - K^{(1)} | \Phi_v \rangle - U_{PV}^{(1)}  |\Phi_v \rangle . \nonumber  
\end{eqnarray}
Here ${\cal E}_v={\cal E}_0 + \epsilon_v$ and $U_{PV}^{(1)}$ is the perturbed DHF potential and is defined as
\begin{eqnarray}
  U_{PV}^{(1)} | \Phi_i \rangle = \sum_b \left [ \langle b | V_{es} \Omega_i^{CPDF,(1)} |b \rangle |i \rangle \right. \nonumber  \\ 
  \left.  -  \langle b | V_{es} \Omega_i^{CPDF,(1)}  |i \rangle |b \rangle \right. \nonumber  \\ 
  \left. + \langle b | \Omega_0^{CPDF,(1)\dagger} V_{es}  | b \rangle |i \rangle \right. \nonumber  \\ 
  \left. -  \langle i |\Omega_0^{CPDF,(1)\dagger} V_{es}  |i \rangle |b \rangle  \right ] .  
\label{eqhfu1c}
\end{eqnarray}
In Eq. (\ref{eqncpdf1}), the first two terms correspond to Core contribution while the last two terms give Valence contributions.
In the RPA method, one can define the wave operators $\Omega_0^{RPA,(1)}$ and $\Omega_i^{RPA,(1)}$ in a similar manner except that $H_{at}$ is replaced by $H_{at}-\omega$ and $K^{(1)}$ by $D$ in the above equations. 

It can be noticed the CPDF and RPA methods still use DHF wave operators to define the unperturbed wave functions that miss out correlation contributions from $V_{es}$. This is partially addressed by the CPDF-RPA method in which the first-order $|\Psi_v^{(1,0)} \rangle$ and $|\Psi_v^{(0,1)} \rangle$ wave functions are determined  the same way as the CPDF and RPA methods, respectively, but there are additional term arising through  $|\Psi_i^{(1,1)} \rangle$ by expressing \cite{Chakraborty2023}
\begin{eqnarray}
{\cal X}_{PV}^{NSD} &=& {\cal C} \langle \Phi_f | \Omega_0^{DHF,(0) \dagger} \overline{D} \Omega_0^{CPDF,(1) } | \Phi_i \rangle \nonumber \\
&& + \langle \Phi_f | \Omega_0^{RPA,(1) \dagger} \overline{K}^{(1)} \Omega_0^{DHF,(0) } | \Phi_i \rangle \nonumber \\
&& + \langle \Phi_f | \Omega_f^{DHF,(0) \dagger} \overline{D} \Omega_i^{CPDF,(1) } | \Phi_i \rangle \nonumber \\
&& + \langle \Phi_f | \Omega_f^{RPA,(1) \dagger}  \overline{K}^{(1)} \Omega_i^{DHF,(0) } | \Phi_i \rangle \nonumber \\
&& + \langle \Phi_f | \Omega_f^{DHF,(0) \dagger} \Omega_i^{(1,1) } | \Phi_i \rangle,
\label{eqncpdf}
\end{eqnarray}
where the CPDF-RPA coupled wave operators are defined as
\begin{eqnarray}
(H_{DHF} - {\cal E}_0- \omega) \Omega_0^{(1,1) } |\Phi_0 \rangle = - D \Omega_0^{CPDF,(1)} | \Phi_i \rangle \nonumber \\  - U_{PV}^{CPDF,(1)}  |\Phi_0 \rangle - K^{(1)}\Omega_i^{RPA,(1)} | \Phi_0 \rangle \nonumber \\ -  U_{PV}^{RPA,(1)}  |\Phi_0 \rangle - U_{PV}^{(1,1)}  |\Phi_0 \rangle  \nonumber  
\end{eqnarray}
and
\begin{eqnarray}
(H_{DHF} - {\cal E}_i- \omega) \Omega_i^{(1,1) } |\Phi_i \rangle =  - D \Omega_i^{CPDF,(1)} | \Phi_i \rangle \nonumber \\ - U_{PV}^{CPDF,(1)}  |\Phi_i \rangle - K^{(1)}\Omega_i^{RPA,(1)} | \Phi_i \rangle \nonumber \\ - U_{PV}^{RPA,(1)}  |\Phi_i \rangle - U_{PV}^{(1,1)}  |\Phi_i \rangle  . \nonumber  
\end{eqnarray}
In this case, we define
\begin{eqnarray}
  U_{PV}^{(1,1)} | \Phi_i \rangle = \sum_b \left [ \langle b | \Omega_i^{RPA,(1)\dagger} V_{es} \Omega_i^{CPDF,(1)} |b \rangle | i \rangle \right. \nonumber \\ \left. -  \langle b | \Omega_i^{RPA,(1)\dagger} V_{es} \Omega_i^{CPDF,(1)}  |i \rangle  |b \rangle \right.  \nonumber  \\ 
  \left. +   \langle  b | \Omega_i^{CPDF,(1)\dagger} V_{es} \Omega_i^{RPA,(1)} | b \rangle  |i \rangle \right. \nonumber \\ \left. -  \langle b | \Omega_i^{CPDF,(1)\dagger} V_{es} \Omega_i^{RPA,(1)}  |i \rangle  |b \rangle \right. \nonumber  \\ 
 \left. + \langle  b | V_{es} \Omega_i^{(1,1)} | b \rangle  |i \rangle -  \langle b  | V_{es} \Omega_i^{(1,1)}  |i \rangle | b \rangle \right.  \nonumber  \\ 
 \left.  + \langle b | \Omega_0^{(1,1)\dagger} V_{es}  |b \rangle  |i \rangle -  \langle b  |\Omega_0^{(1,1)\dagger} V_{es}  |i  \rangle |b \rangle \right ] . 
\label{eqhfu1}
\end{eqnarray}
As can be seen from the above expression, there are wave operators from both the CPDF and RPA methods are coupled through $V_{es}$ in the CPDF-RPA method. This is how additional core-polarization effects as well as DCP effects are arising in the CPDF-RPA method. Due to complexity involved in the evaluation of DCP terms and their contributions are being too small, sometime they are neglected in the studies of PV effects in the atomic systems (refer to Ref. \cite{Chakraborty2023} for more discussions on this). To demonstrate importance of the DCP contributions to ${\cal X}_{PV}^{NSD}$, we present results without accounting and after including the DCP contributions in the CPDF-RPA method. The CPDF-RPA method without the DCP effects is denoted by CPDF-RPA$^*$ method in this work.    

Though the CPDF, RPA, and CPDF-RPA methods are all-order methods, they only take into account CP effects to all-orders through single excitations. To account for the effects of both PC and CP on all-orders, along with their correlations, it is necessary to consider doubly excited state configurations. In view of this consideration, RCC theory is a better choice. In the RCC theory {\it ansatz}, the unperturbed wave operators are given by \cite{Chakraborty2023}
\begin{eqnarray}
\Omega_0^{(0)} = e^{T^{(0)}} |\Phi_0 \rangle 
\end{eqnarray}
and
\begin{eqnarray}
\Omega_v^{(0)} = e^{T^{(0)}} S_v^{(0)} |\Phi_v \rangle .
\label{eqrcc}
\end{eqnarray}
Extending these definitions to the first-order perturbed wave functions, we can define the corresponding wave operators as \cite{Chakraborty2023}
\begin{eqnarray}
\Omega_0^{(1)} = e^{T^{(0)}} T^{(1)} |\Phi_0 \rangle 
\end{eqnarray}
and
\begin{eqnarray}
\Omega_v^{(1)} = e^{T^{(0)}} \left ( S_v^{(1)} + (1+S_v^{(0)})T^{(1)} \right )  |\Phi_v \rangle .
\end{eqnarray}

The amplitude determining equations for the unperturbed wave operator in the (R)CC theory are given by \cite{Mukherjee1979, Lindgren1985}
\begin{eqnarray}
\langle \Phi_a^p | [H_0,\Omega_0^{(0)}] |\Phi_0 \rangle= - \langle \Phi_a^p | V_{es} \Omega_0^{(0)}  |\Phi_0 \rangle
\end{eqnarray}
and
\begin{eqnarray}
\langle \Phi_v^p | [H_0,\Omega_v^{(0)}] |\Phi_v \rangle &=& - \langle \Phi_v^p | V_{es} (\Omega_0^{(0)} + \Omega_v^{(0)}) |\Phi_v \rangle \nonumber \\
 &+&  E^{(0)}_v \langle \Phi_v^p | \Omega_v^{(0)}  |\Phi_v \rangle .
\end{eqnarray}
 The energy, $E^{(0)}_v$, of the state is estimated by
 \begin{eqnarray}
E^{(0)}_v=\langle \Phi_v | H_{at} (\Omega_0^{(0)} +\Omega_v^{(0)}) |\Phi_v \rangle .
 \end{eqnarray}

 After the amplitudes of the unperturbed wave operators are known, it is also possible to evaluate the E1 matrix elements and matrix elements of the $K^{(1)}$ operator in the RCC theory using the following expression  
\begin{eqnarray}
{\cal O} = \frac{\langle \Phi_f | \{1+S_f^{(0)} \}^{\dagger}  \bar{O} \{ 1+ S_i^{(0)} \} |\Phi_i \rangle} {\langle \Phi_f | \{1+ S_f^{(0)} \}^{\dagger} \bar{N} \{ 1+ S_i^{(0)} \} |\Phi_i \rangle} , 
\label{e1mat}
\end{eqnarray}
where $O$ denotes either the $D$ or $K^{(1)}$ operator, $\bar{O}=e^{T^{(0)\dagger}}Oe^{T^{(0)}}$ and $\bar{N}=e^{T^{(0)\dagger}}e^{T^{(0)}}$. By using these matrix elements and energies in either Eq. (\ref{eq4}) or Eq. (\ref{eq6}), we can evaluate the ${\cal X}_{PV}^{NSD}$ values as a sum-over-states approach. In fact by comparing the calculated energies and E1 matrix elements with their experimental results, accuracy of the calculated ${\cal X}_{PV}^{NSD}$ values can be gauged. It is also possible to replace the calculated energies and E1 matrix elements in a sum-over-states approach by their experimental values to improve accuracy of the ${\cal X}_{PV}^{NSD}$ values by a semi-empirical treatment of the method. Nonetheless, the sum-over-states approach has the limitation that it can only include matrix elements involving a few low-lying intermediate states (denoted as ``Main" contribution) that are bound and have singly excited configurations with respect to the initial and final DHF states. Since PV interaction originates from the nucleus and densities of continuum over the nucleus are large in comparison to the high-lying bound states, it is important to include contributions from the continuum to achieve accurate calculations of the ${\cal X}_{PV}^{NSD}$ values. We referred to the contributions from the continuum and high-lying bound states as ``Tail". Also, contributions from the doubly excited configurations, which are a part DCP effects, cannot be included in the sum-over-states approach. These two contributions can be included more rigorously by solving equations directly for the first-order perturbed wave functions than expressing them in terms of sum over intermediate states.    
 
To obtain the first-order perturbed wave functions in the first-principle approach using the RCC theory, amplitudes of the first-order perturbed wave operators can be obtained using the following equations \cite{Sahoo2021, Sahoo2006} 
\begin{eqnarray}
\langle \Phi_a^p | [H_0,\Omega_0^{(1)}] |\Phi_0 \rangle= - \langle \Phi_a^p | K^{(1)} \Omega_0^{(0)} + V_{es} \Omega_0^{(1)}  |\Phi_0 \rangle \ \ \ \
\end{eqnarray}
and
\begin{eqnarray}
\langle \Phi_v^p | [H_0,\Omega_v^{(1)}] |\Phi_v \rangle &=& - \langle \Phi_v^p | K^{(1)} (\Omega_0^{(0)} + \Omega_v^{(0)}) |\Phi_v \rangle \nonumber \\
&& + \langle \Phi_v^p | V_{es} (\Omega_0^{(1)} + \Omega_v^{(1)}) |\Phi_v \rangle \nonumber \\
 && +  E^{(0)}_v \langle \Phi_v^p | \Omega_v^{(1)} |\Phi_v \rangle .
\end{eqnarray} 

 In this work,  we use Eq. (\ref{eq3}) to evaluate ${\cal X}_{PV}^{NSD}$ which follows
\begin{eqnarray}
{\cal X}_{PV}^{NSD} = {\cal C} \frac{\langle \Phi_f | \{S_f^{(1)\dagger} + (1+S_f^{(0)\dagger}) T^{(1)\dagger}\} \bar{D} \{ 1+ S_i^{(0)} \} |\Phi_i \rangle} {\langle \Phi_f | \{S_f^{(0)\dagger} +1 \} \bar{N} \{ 1+ S_i^{(0)} \} |\Phi_i \rangle}  \nonumber \\
 +  \frac{\langle \Phi_f |\{ 1+ S_f^{(0)} \}^{\dagger} \bar{D} \{T^{(1)}(1+ S_i^{(0)}) + S_i^{(1)}\} |\Phi_i \rangle}{\langle \Phi_f | \{S_f^{(0)\dagger} +1 \} \bar{N} \{ 1+ S_i^{(0)} \} |\Phi_i \rangle} . \ \ \ \ \ \ \ 
\label{e1pnc}
\end{eqnarray}
From the above expression, contributions arising through $\bar{D} T^{(1)}$ and $T^{(1)\dagger} \bar{D}$ correspond to Core contribution while contributions from the rest of the terms are the Valence contributions.

In the present work, we consider all possible single and double excitations in the RCC theory (RCCSD method). The singles and doubles excited RCC wave operators are denoted by additional subscripts 1 and 2, respectively. Thus, we define in the RCCSD method
\begin{eqnarray}
T_0^{(0)} &=&  T_{10}^{(0)} + T_{20}^{(0)} , \nonumber \\
T_0^{(1)} &=&  T_{10}^{(1)} + T_{20}^{(1)}, \nonumber \\
S_v^{(0)} &=&  S_{1v}^{(0)} + S_{2v}^{(0)} \nonumber 
\end{eqnarray}
and
\begin{eqnarray}
S_v^{(1)} &=&  S_{1v}^{(1)} + S_{2v}^{(1)} .
\end{eqnarray}

As mentioned earlier, the DCP contributions are included in the CPDF-RPA method due to the last term of Eq. (\ref{eq5}). We had a choice to adopt Eq. (\ref{eq5}) in the RCC theory to determine ${\cal X}_{PV}^{NSD}$. However, such an approach would demand to compute and store amplitudes of the $\Omega_0^{(1,1)}$ and $\Omega_v^{(1,1)}$ wave operators. This is a quite bit of involved work and requires a large computational resource to carry out the calculations. In contrast to the other methods, electron correlation effects are included in the evaluation of both the bra and ket states through the RCC theory. As a result, it includes a lot more electron correlation effects compared to the CPDF, RPA and CPDF-RPA methods. Following discussions in Ref. \cite{Chakraborty2023}, it can be shown that the RCCSD term $D S_{2v}^{(1)}$ and its complex conjugate (c.c) include all DCP contributions of the CPDF-RPA method even though Eq. (\ref{e1pnc}) is derived based on Eq. (\ref{eq3}). This is the potential of the RCC theory for which the results obtained using the RCCSD method can be treated as more accurate compared to the other methods.  

\begin{table}[t!]
\caption{The Core and Valence contributions to the ${\cal X}_{PV}^{NSD}$ values in all the considered transitions among all possible hyperfine levels $F_f$ and $F_i$ of the $7s ~ ^2S_{1/2}$ and $6s ~ ^2S_{1/2}$ states, respectively, in $^{133}$Cs from different methods. Same units as in Table \ref{tab0} are used here.} 
\centering
\begin{tabular}
{p{2.5cm}p{0.5cm}p{0.5cm}p{0.3cm}p{2.2cm}p{2.0cm}}
\hline\hline
Method & $F_f$ & $F_i$ & & Core & Valence\\
\hline \\
DHF & 3 & 3 && $-0.0046$ & 1.9075 \\
    & 3 & 4 && $-0.2031$ & 5.6693  \\
    & 4 & 3 && $-0.2014$ & 4.9350 \\
    & 4 & 4 && $-0.0051$ & 2.1717 \\
  & &  &  & & \\   
CPDF& 3 & 3 && $-0.0049$ & 2.3394  \\
    & 3 & 4 && $-0.4417$ & 7.4872 \\
    & 4 & 3 && $-0.4396$ & 6.5866 \\
    & 4 & 4 && $-0.0056$ & 2.6635 \\
   & &  &  & & \\   
RPA& 3 & 3 && ~~0.0007 & 1.8298  \\
   & 3 & 4 && $-0.2814$ & 5.9552 \\
   & 4 & 3 && $-0.2821$ & 5.2510 \\
   & 4 & 4 && ~~0.0007 & 2.0835 \\
   & &  &  & & \\  
CPDF-RPA*& 3 & 3 && ~~0.0039 & 2.2417  \\
        & 3 & 4 && $-0.6181$ & 7.8529  \\
        & 4 & 3 && $-0.6195$ & 6.9902  \\
        & 4 & 4 && ~~0.0039 & 2.5525 \\
     & & &  &  & \\     
CPDF-RPA& 3 & 3 && ~~0.0039 & 2.0100 \\
        & 3 & 4 && $-0.6181$ & 7.6072  \\
        & 4 & 3 && $-0.6195$ & 6.8337  \\
        & 4 & 4 && ~~0.0039 & 2.2889\\
   & & &  &  & \\       
RCCSD& 3 & 3 && $-0.0047$ & 2.3392  \\
     & 3 & 4 && $-0.3458$ & 7.7401 \\
     & 4 & 3 && $-0.3441$ & 6.8399 \\
     & 4 & 4 && $-0.0052$ & 2.6627 \\[0.6ex]
\hline\hline
\end{tabular}
\label{tab1}
\end{table}

\section{Results \& Discussion}

We present our calculated ${\cal X}_{PV}^{NSD}$ values between the $F=3$ and $F=4$ hyperfine levels of the $6s ~ ^2S_{1/2} \rightarrow 7s ~ ^2S_{1/2}$ transition in the $^{133}$Cs atom in Table \ref{tab0}. They are given from the DHF, CPDF, RPA, CPDF-RPA*, CPDF-RPA and RCCSD methods. In the same table, we have also given results that were reported earlier in the literature \cite{Johnson2003, Safronova2009, Mani2011}. The differences in the results between the DHF and other methods listed in the table show amount of electron correlation included through the respective method. Analyses of these results show a very interesting trend. By comparing results between the CPDF and RPA method, we can draw the conclusion that CP effects arising through the $K^{(1)}$ much stronger than those arise through the E1 operator. In fact, they contribute with opposite signs with respect to the DHF values. Interestingly this trend is more peculiar in the CPDF-RPA* method, in which results between the $F_i=3\rightarrow F_f=4$ and $F_i=4 \rightarrow F_f=3$ hyperfine levels are larger than the CPDF values while they are smaller for the $F_i=3\rightarrow F_f=3$ and $F_i=4\rightarrow F_f=4$ hyperfine levels compared to the CPDF values. So from these results it would not be clearly argued that the CP effects through the E1 operator always contribute with an opposite sign than the $K^{(1)}$ operator. The differences between the results from the CPDF-RPA* and CPDF-RPA indicate that the DCP contributions are quite significant in the evaluations of the ${\cal X}_{PV}^{NSD}$ values and they reduce the values obtained using the CPDF-RPA* method. The results from the RCCSD method are seen to be larger than the CPDF-RPA method. The RCCSD method includes all contributions that are taken into account in the CPDF-RPA method as well as contributions from the PC effects and correlations among both the CP and PC effects to all-orders. Though the CPDF-RPA* values appear to be close to the RCCSD values, both the CPDF-RPA and RCCSD methods include DCP contributions, while the CPDF-RPA* method does not. It, thus, suggests that contributions arising through the PC effects in the determination of the ${\cal X}_{PV}^{NSD}$ values are quite significant and cannot be neglected for their accurate evaluation. To gauge the magnitudes of the ${\cal X}_{PV}^{NSD}$ values quantitatively for the transitions between different hyperfine levels, we plot these values in Fig. \ref{fig0}. This clearly shows that magnitudes of the ${\cal X}_{PV}^{NSD}$ values in the $F_i=3 \rightarrow F_f=3$ and $F_i=4 \rightarrow F_f=4$ transitions are smaller compared to the $F_i=3 \rightarrow F_f=4$ and $F_i=4 \rightarrow F_f=3$ transitions.

As can be seen in Table \ref{tab0}, calculations carried out at the DHF level in Ref. \cite{Johnson2003} agree quite well with our values but differ a little bit from Ref. \cite{Mani2011}. A careful comparison shows, our CPDF-RPA* results match well with the RPA values of Ref. \cite{Johnson2003}. From this analysis, we assume that the RPA method of Ref. \cite{Johnson2003} is not exactly same with our RPA method rather it considers the combined results from both the CPDF and RPA methods. It, however, appears to us that DCP contributions were not included in Ref. \cite{Johnson2003}. This can also be corroborated with the expression used in Ref. \cite{Johnson2003}, i.e. Eq. (\ref{eq4}), in the determination of ${\cal X}_{PV}^{NSD}$ as this expression is difficult to implement in the form of Eq. (\ref{eqncpdf}). As stated categorically earlier, we are able to include the DCP contribution in the CPDF-RPA method by using the formula given by Eq. (\ref{eq6}). The calculations in Ref. \cite{Safronova2009} differ from the present work on two major grounds. First, it uses a sum-over-states approach in which E1 and $K^{(1)}$ matrix elements of a few low-lying transitions are explicitly evaluated using the RCC theory and experimental energy values were used. Limitations of the sum-over-states approach has already been discussed earlier. Second, only linear in $T^{(0)}$ and $S_v^{(0)}$ terms appearing in Eq. (\ref{eqrcc}) of the singles and doubles approximated RCC theory (SD) are being considered in the evaluation of the matrix elements. The differences between the sum-over-states SD and {\it ab initio} RCCSD results could be due to the DCP effects and contributions from the Core and Tail contributions, which are included more rigorously in the present work. From the comparison between the results from the PRCC method of Ref. \cite{Mani2011} and RCCSD method of present work shows large differences. This could be due to two reasons; use of different size basis functions and difference in the implementation of the method. In both the works, though Gaussian type orbitals (GTOs) are being used, different set of parameters are being considered. Since we have compared our DHF values with those of Ref. \cite{Johnson2003} and the results are agreeing, we presume that our basis functions are good enough to produce accurate results for ${\cal X}_{PV}^{NSD}$. It should be noted that B-spline polynomials were used as basis functions in Ref. \cite{Johnson2003} in contrast to our GTOs. From the point of view of method, both the PRCC and RCCSD methods are the singles and doubles approximated RCC theory, and obtain results in the first principle approach. However, their implementation procedures in Ref. \cite{Mani2011} and in our work differ in dealing with the angular momentum couplings between the nuclear and electronic components. We follow the procedures adopted by other groups to decouple nuclear and electronic angular factors, and carry out calculations only on the electronic coordinate by expressing nuclear angular factors as prefactors. In contrast, both the nuclear and electronic couplings are included together in the PRCC method. Nonetheless, the results are expected to be similar in both approaches as nuclear and electronic components of the wave functions are treated at the same level of approximation. To ensure correct implementation of our method, we reproduced the results of the CPDF-RPA method by the corresponding terms of the RCCSD method and extra contributions arising through the RCCSD method over the CPDF-RPA method have been investigated thoroughly. From this analysis, we presume that our RCCSD calculations are more reliable.      

\begin{figure}[t!]
\centering
\begin{tabular}{c c c}\\
\includegraphics[width=40mm,height=40mm]{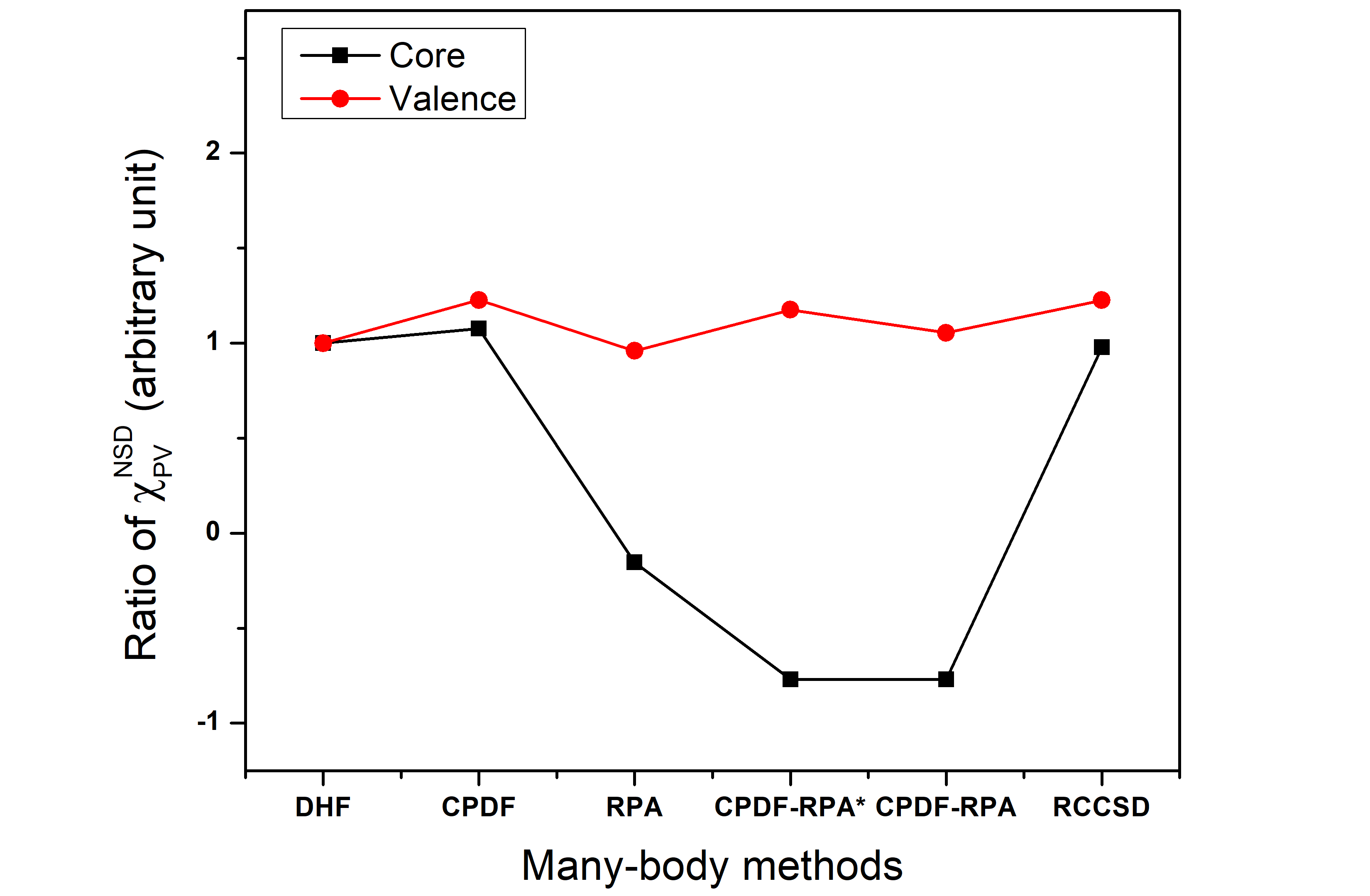} & &
\includegraphics[width=40mm,height=40mm]{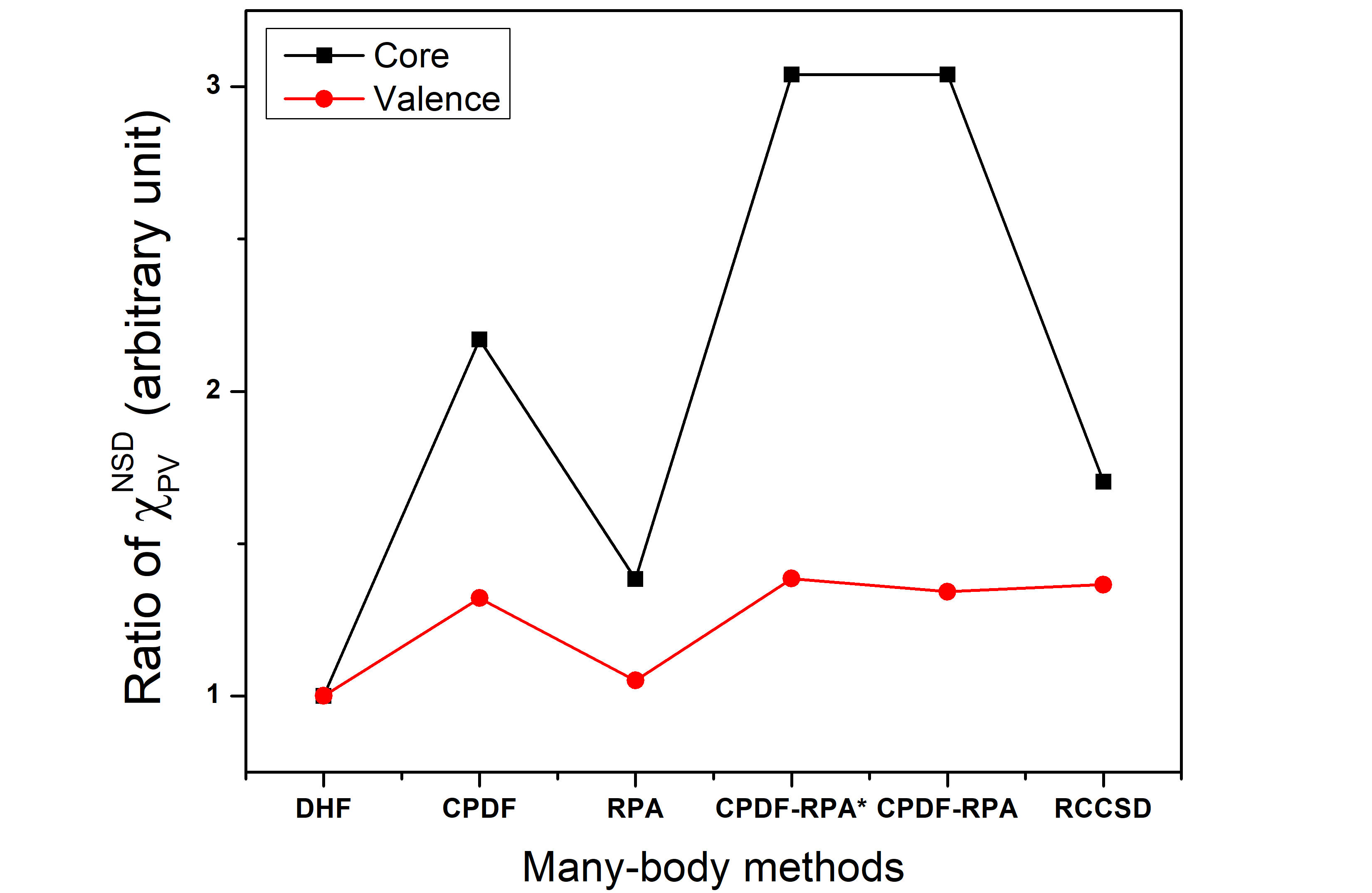}\\ 
(a) $F_f=3 \rightarrow F_i=3$ & & (b) $F_f=3 \rightarrow F_i=4$  \\
  &  & \\
\includegraphics[width=40mm,height=40mm]{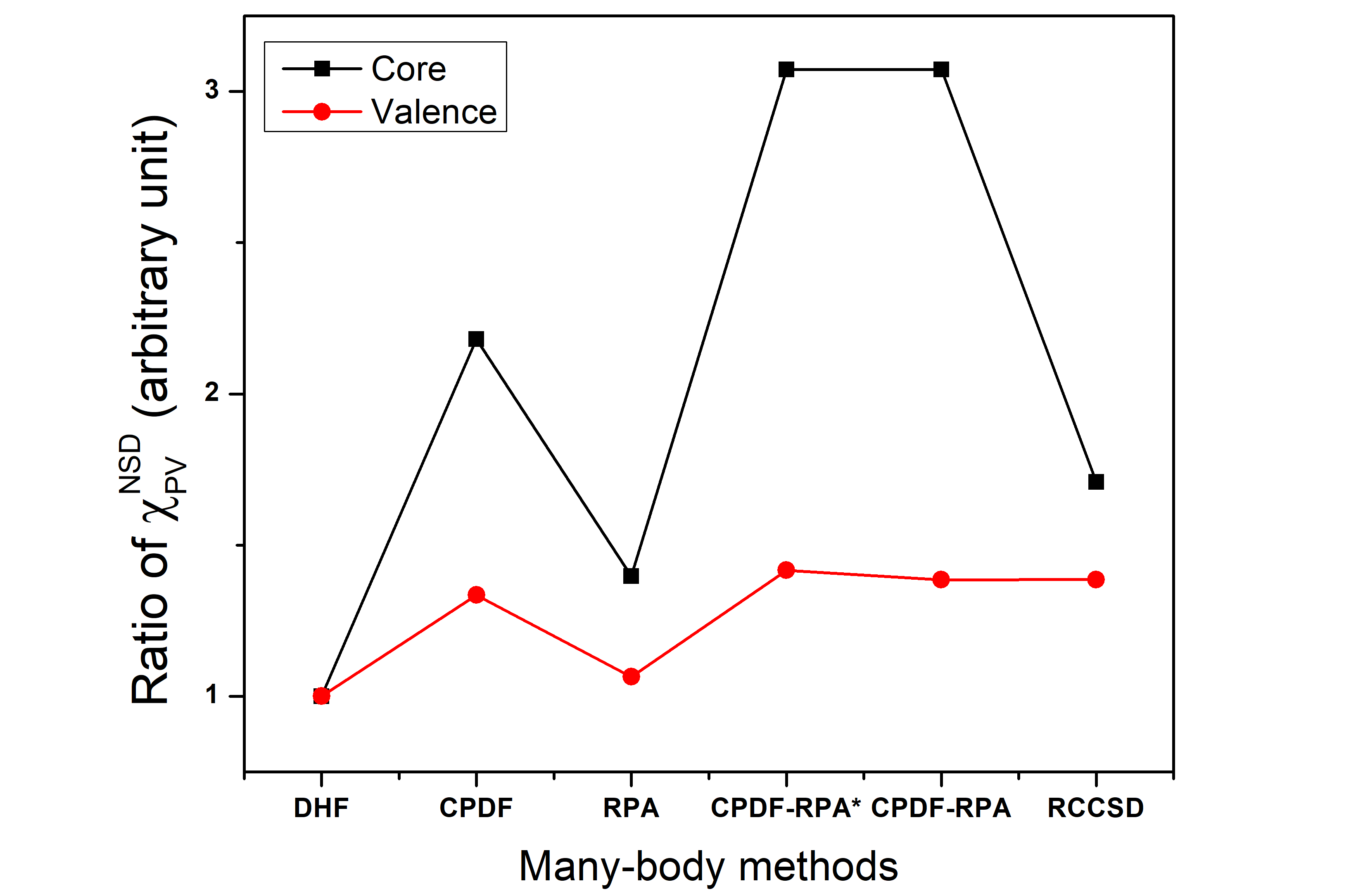} & &
\includegraphics[width=40mm,height=40mm]{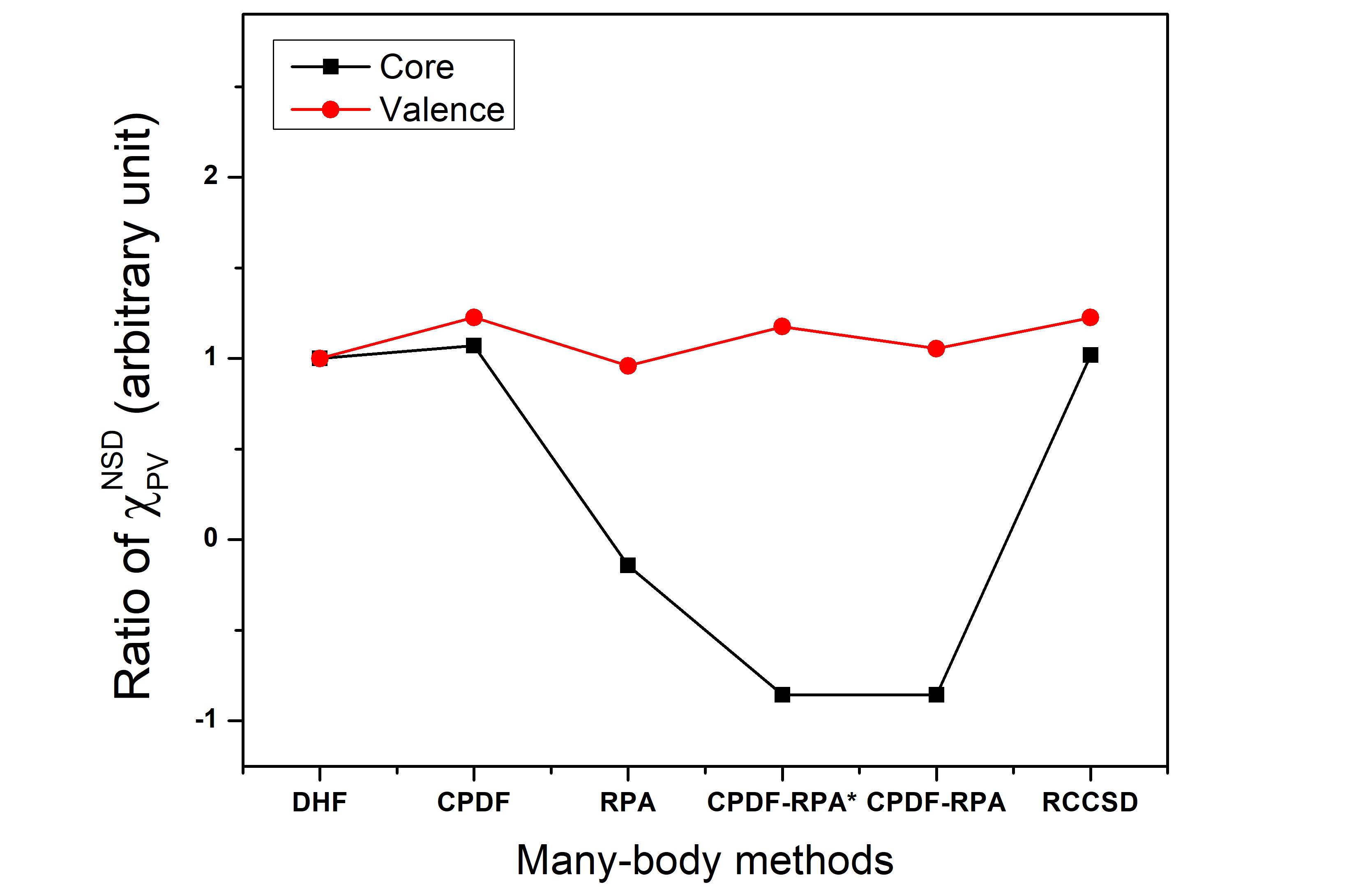}\\ 
(c) $F_f=4 \rightarrow F_i=3$ & & (d) $F_f=4 \rightarrow F_i=4$  \\
\end{tabular}
\caption{Ratios of the calculated ${\cal X}_{PV}^{NSD}$ values for the $F_f-F_i$ transitions of the $7s ~ ^2S_{1/2}$ and $6s ~ ^2S_{1/2}$ states, respectively, in $^{133}$Cs from different methods with respect to their DHF values. Same units as in Table \ref{tab0} are used}
\label{fig1}
\end{figure}

We also try to understand the roles of both the Core and Valence correlation contributions in the evaluation of ${\cal X}_{PV}^{NSD}$. For this purpose, we present both these contributions from each considered method in Table \ref{tab1}. Comparison of Core and Valence contributions from each method for different hyperfine transitions shows how both types of effects vary from lower to higher-order approximations in the many-body theory. The first interesting phenomena one can observe from this table is that the Core contributions to the $F_i=3 \rightarrow F_f=3$ and $F_i=4 \rightarrow F_f=4$ transitions are similar but the Valence contributions to these transitions differ significantly. An analogous behavior can also be noticed for the $F_i=3 \rightarrow F_f=4$ and $F_i=4 \rightarrow F_f=3$ transitions. It means that the similarity between Core contributions to these transitions cannot be attributed to their radial and angular factors, but it is a peculiar behavior of the property being studied. Another difference in the trend can be noticed that the signs of the Core contributions among the DHF, CPDF and RCCSD are opposite for the $F_i=3 \rightarrow F_f=3$ and $F_i=4 \rightarrow F_f=4$ transitions than the RPA, CPDF-RPA* and CPDF-RPA methods, while they are same for the $F_i=3 \rightarrow F_f=4$ and $F_i=4 \rightarrow F_f=3$ transitions. It also suggests that the Core contributions arising from the combined CPDF-RPA (or CPDF-RPA*) method are not same as the total sum of the individual Core contributions from both the CPDF and RPA methods. Thus, the correlation trends in these methods behave completely different from each other. It is worth mentioning here that in the CPDF and RCCSD method $K^{(1)}$ is treated as perturbation. If the E1 operator is treated perturbatively in the RCCSD method like in the RPA, the sign of Core contribution may differ. However, results at the DHF method is independent of whether $K^{(1)}$ or E1 operator is considered as perturbation. Now from the analyses of Valence contributions, we do not find any sign differences among their values for any of the hyperfine transition at different levels of approximation in the methods, but their magnitudes show large variance from one to another method. The differences between the values from the DHF method and other methods for the $F_i=3 \rightarrow F_f=3$ and $F_i=4 \rightarrow F_f=4$ transitions are relatively smaller than the $F_i=3 \rightarrow F_f=4$ and $F_i=4 \rightarrow F_f=3$ transitions.

\begin{table}[t!]
\caption{Comparison of contributions from the initial and final perturbed states to ${\cal X}_{PV}^{NSD}$ among all possible hyperfine levels of the $7s ~ ^2S_{1/2} (F_f)-6s ~ ^2S_{1/2}(F_i)$ transition in $^{133}$Cs at different methods, in units of $i e a_0 K_W \times10^{-12}$.}
\begin{ruledtabular}
\scalebox{0.95}{\begin{tabular}{l c c c c c}
Method & $F_f$ & $F_i$  & $\langle7S^{(0)}|\tilde{D}_f|6S^{(1)}\rangle$ & $\langle 7S^{(1)}|\tilde{D}_i|6S^{(0)}\rangle$ & Total \\
\hline \\
DHF & 3 & 3 & $-0.7077$ & 2.6106 & 1.9029 \\
    & 3 & 4 &  ~~0.9513 & 4.5150 & 5.4663 \\
    & 4 & 3 &  ~~1.2236 & 3.5101 & 4.7337 \\
    & 4 & 4 & $-0.8057$ & 2.9722 & 2.1665 \\
 & & &  &  & \\
CPDF & 3 & 3 & $-0.8673$ & 3.2018 & 2.3345 \\
     & 3 & 4 & ~~1.2201 & 5.8254 & 7.0455 \\
     & 4 & 3 & ~~1.5540 & 4.5930 & 6.1470 \\
     & 4 & 4 & $-0.9873$ & 3.6452 & 2.6579 \\
 & & &   & & \\
RPA & 3 & 3 & $-0.8158$ & 2.6463 & 1.8305 \\
    & 3 & 4 & ~~1.0969 & 4.5769 & 5.6738 \\
    & 4 & 3 & ~~1.4108 & 3.5581 & 4.9689 \\
    & 4 & 4 & $-0.9289$ & 3.0131 & 2.0842 \\
 & & & & &  \\
CPDF-RPA*& 3 & 3 & $-0.9901$ & 3.2357 & 2.2456 \\
        & 3 & 4 & ~~1.3688 & 5.8660 & 7.2348 \\
        & 4 & 3 & ~~1.7504 & 4.6203 & 6.3707 \\
        & 4 & 4 & $-1.1277$ & 3.6841 & 2.5564 \\
 & & & &  & \\
CPDF-RPA& 3 & 3 & $-0.9677$ & 2.9816 & 2.0139 \\
        & 3 & 4 & ~~1.3163 & 5.6728 & 6.9891 \\
        & 4 & 3 & ~~1.6894 & 4.5248 & 6.2142 \\
        & 4 & 4 & $-1.1018$ & 3.3946 & 2.2928 \\
 & & & &  &  \\
RCCSD & 3 & 3 & $-0.9807$ & 3.3151 & 2.3344 \\
      & 3 & 4 & ~~1.4815 & 5.9128 & 7.3943 \\
      & 4 & 3 & ~~1.8590 & 4.6367 & 6.4958 \\
      & 4 & 4 & $-1.1169$ & 3.7744 & 2.6575 \\[0.6ex]
\end{tabular}}
\end{ruledtabular}
\label{tab2}
\end{table}

To fathom the role of electron correlation effects accounted through each many-body method in a more quantified view point, we plot the ${\cal X}_{PV}^{NSD}$ values from each method with respect to the DHF values (by taking ratio of each calculation with respect to the corresponding DHF value) for each transition between the hyperfine levels in Fig. \ref{fig1}. As the figure clearly shows, the correlation trends are different in the estimation of Core and Valence contributions at different approximations of the method. Both the net magnitude and correlation effects to Core contribution are small while they are pronounced in the estimation of Valence contribution. This is the reason why earlier reported results using a sum-over-states approach, in which only Valence contributions are estimated more rigorously, look reasonably accurate. Further looking closely at the figure, it can be said that the correlation effects included through the CPDF-RPA method are the largest compared to the other methods. In the $F_i=3 \rightarrow F_f=3$ and $F_i=4 \rightarrow F_f=4$ transitions, the CPDF method includes the smallest amount of correlation effects while in the $F_i=3 \rightarrow F_f=4$ and $F_i=4 \rightarrow F_f=3$ transitions the RPA method accounts for the smallest amount of correlation effects. Interestingly the correlation effects arising through the RCCSD method are next to the second smallest in all the transitions though it includes all effects that are included through the CPDF, RPA, and CPDF-RPA methods. It, therefore, suggests that the PC effects that are included through the RCCSD method and missing in the other methods contribute almost equally, but with opposite signs.    

\begin{table}[t!]
\caption{Contributions from different RCC terms to the ${\cal X}_{PV}^{NSD}$ values (in units $i e a_0 K_W \times10^{-12}$) of the $7s ~ ^2S_{1/2} (F_f) - 6s ~ ^2S_{1/2}(F_i)$ transitions in $^{133}$Cs. Here, contributions given under `Norm' represent the corrections to results due to normalization factors of the wave functions. In this table $\overline{D}$ denotes only the effective one-body part of $e^{T^{(0)\dagger}}De^{T^{(0)}}$. Contributions from other non-linear terms of the RCCSD method are given together under ``Others".}
\begin{ruledtabular}
\begin{tabular}{lrrrr}
  RCC term   &  $3-3$ & $3-4$ & $4-3$  & $4-4$ \\
\hline \\

 $\bar{D}T_1^{(1)}$  & $-0.1029$ & $-0.1981$ & $-0.1585$ & $-0.1169$\\
 
 $T_1^{(1)\dagger} \bar{D}$  & 0.0983 & $-0.1501$ & $-0.1879$  & 0.1120 \\
 
 $\bar{D}S_{1i}^{(1)}$  & $-0.4487$ & $0.6832$ & 0.8561  & $-0.5110$ \\
 
 $S_{1f}^{(1)\dagger} \bar{D}$  & 4.5384 & 8.0486 & 6.3017  & 5.1674 \\
 
 $S_{1f}^{(0)\dagger} \bar{D} S_{1i}^{(1)}$ & $-0.5575$ & 0.8032 & 1.0178  & $-0.6349$ \\
 
 $S_{1f}^{(1)\dagger} \bar{D} S_{1i}^{(0)}$  & $-1.0339$  & $-1.8588$ & $-1.4609$ & $-1.1770$ \\
 
 $\bar{D}S_{2i}^{(1)}$  & $-0.0577$ & 0.1179 & 0.1403 & $-0.0658$ \\
 
 $S_{2f}^{(1)\dagger} \bar{D}$ & $-0.0077$ & 0.0273 & 0.0304  & $-0.0087$ \\
 
 Others  & $-0.0389$ & 0.0947 & 0.1090  & $-0.0452$ \\
 Norm&  $-0.0549$ & $-0.1736$ & $-0.1522$ & $-0.0624$ \\
 
 \hline\\
 Total  & 2.3344 & 7.3943 & 6.4958  & 2.6575 \\
 
\end{tabular}
\end{ruledtabular}
\label{tab3}
\end{table} 

Now we intend to understand contributions to the ${\cal X}_{PV}^{NSD}$ values arising through the initial and final perturbed states in different methods. In Table \ref{tab2} we have presented results from both the initial and final perturbed states separately in all the hyperfine level transitions using the methods that are employed in this work. As can be seen from this table, the final perturbed states contribute predominantly over the initial perturbed states among all the transitions at different approximation of methods irrespective of whether the $K^{(1)}$ or E1 operator is considered as perturbation. This trend has striking similarity with the evaluation of the $E1_{PV}^{NSI}$ amplitude of the $6s ~ ^2S_{1/2}-7s ~ ^2S_{1/2}$ transition in $^{133}$Cs \cite{Chakraborty2023}. Another observation is that contributions to ${\cal X}_{PV}^{NSD}$ from the initial and final states in the $F_i=3 \rightarrow F_f=3$ and $F_i=4 \rightarrow F_f=4$ transitions are in opposite sign, while they contribute with same sign in the $F_i=3 \rightarrow F_f=4$ and $F_i=4 \rightarrow F_f=3$ transitions resulting in enhancement in the final result. 

\begin{table}[t!]
\centering
\caption{Demonstration of variations in the RCCSD values of ${\cal X}_{PV}^{NSD}$ among different hyperfine levels of the $7s ~ ^2S_{1/2}$ and $6s ~ ^2S_{1/2}$ states with various nuclear charge density distributions. Same units as in Table \ref{tab0} are used here.}
\begin{tabular}{p{0.5cm}p{0.5cm}p{0.2cm}p{1.6cm}p{1.6cm}p{1.6cm}}
\hline\hline
$F_f$ & $F_i$ && \multicolumn{3}{c}{Distribution type}\\
\cline{4-6}
 & && Fermi & Gaussian & Uniform\\
 \hline
 3 & 4 && 2.3344  & 2.3485  & 2.2964 \\
 3 & 4 && 7.3943  & 7.4318  & 7.2620\\
 4 & 3 && 6.4958  & 6.5278  & 6.3781\\
 4 & 4 && 2.6575  & 2.6739  & 2.6145\\
\hline\hline 
\end{tabular}
\label{tab_nuc}
\end{table}

\begin{table*}[t!]
\centering
\caption{Calculated reduced E1 matrix elements (in a.u.), reduced $K^{(1)}$ matrix elements (in $i K_W \times 10^{-12}$) and the excitation energies (in $cm^{-1}$) of the low-lying states of $^{133}$Cs using the RCCSD method. These values have been used to estimate the `Main' contributions to the ${\cal X}_{PV}^{NSD}$ amplitudes. Our results are also compared with the precise experimental values wherever they are available. Note that signs of the experimental E1 matrix elements cannot be determined.}
\begin{ruledtabular}
\begin{tabular}{c rrccc}
Transition & \multicolumn{2}{c}{E1 matrix elements} & $K^{(1)}$ amplitudes &  \multicolumn{2}{c}{Excitation energies}  \\
\cline{2-3}\cline{5-6} \\
& This work & Experiment & &  This work & Experiment \cite{NIST}\\
\hline
$6P_{1/2}-6S_{1/2}$ & 4.5512 & 4.5012(26) \cite{Amiot2002} & $-2.0914$  & 11224.82 & 11178.27 \\

$7P_{1/2}-6S_{1/2}$ & 0.3010 & 0.27810(45) \cite{Damitz2019} & $-1.1801$  & 21809.27 & 21765.35\\

$8P_{1/2}-6S_{1/2}$ & 0.0916 & 0.0723(44) \cite{Morton2000} & $-0.7930$  & 25755.02 & 25708.83\\

$9P_{1/2}-6S_{1/2}$ & $-0.0389$ & & ~~0.5717  & 27702.51 & 27637.00 \\

$6P_{3/2}-6S_{1/2}$ & 6.4009 & 6.3403(64) \cite{Young1994} & ~~0.0370  & 11785.84 & 11732.31 \\

$7P_{3/2}-6S_{1/2}$ & 0.6097 & 0.57417(57) \cite{Damitz2019} & ~~0.0213  & 21992.53 & 21946.40 \\

$8P_{3/2}-6S_{1/2}$ & 0.2326 & & ~~0.0141  & 25838.57 & 25791.51 \\

$9P_{3/2}-6S_{1/2}$ & 0.1243 & & ~~0.0101  & 27745.82 & 27681.68 \\

$7S_{1/2}-6P_{1/2}$ & $-4.2535$ & 4.249(4) \cite{Toh2019} & ~~1.0348  & 7344.17 & 7357.26 \\

$7S_{1/2}-7P_{1/2}$ & 10.3017 & 10.325(5) \cite{Toh2019-2} & ~~0.5837  & 3240.28 & 3229.82 \\

$7S_{1/2}-8P_{1/2}$ & 0.9500 & & ~~0.3917  & 7186.03 & 7173.31 \\

$7S_{1/2}-9P_{1/2}$ & $-0.3870$ & & $-0.2823$  & 9133.52 & 9101.47 \\

$7S_{1/2}-6P_{3/2}$ & 6.5053 & 6.489(5) \cite{Toh2019} & ~~0.0167  & 6783.15 & 6803.22 \\

$7S_{1/2}-7P_{3/2}$ & $-14.3023$ &   14.344(7) \cite{Toh2019-2} & ~~0.0094  & 3423.54 & 3410.87 \\

$7S_{1/2}-8P_{3/2}$ & $-1.6676$ & & ~~0.0067  & 7269.58 & 7255.98 \\

$7S_{1/2}-9P_{3/2}$ & $-0.7372$ & & ~~0.0050  & 9176.83 & 9146.15 \\

\end{tabular}
\end{ruledtabular} 
\label{tab_elemnt}
\end{table*}

\begin{table}[t!]
\centering
\caption{Estimated `Main' Contributions to the ${\cal X}_{PV}^{NSD}$ values, in units of $i e a_0 K_W \times10^{-12}$, of the $7s ~ ^2S_{1/2} (F_f) - 6s ~ ^2S_{1/2} (F_i)$ transitions in $^{133}$Cs using matrix elements involving the $np ~ ^2P_{1/2,3/2}$ ($n=6-9$) intermediate states from the RCCSD method in the sum-over-states approach. The values given by {\it ab initio} results. These values obtained after replacing some of the calculated E1 matrix elements and energies by their precisely known experimental values are given under semi-empirical results. These values are estimated for the initial perturbed and final perturbed states separately, then the final values are given after adding both the contributions.}
\begin{ruledtabular}
\begin{tabular}{c  c  c c  c}
$F_f$ & $F_i$ & $\langle7S^{(0)}|\tilde{D}_f|6S^{(1)}\rangle$ & $\langle 7S^{(1)}|\tilde{D}_i|6S^{(0)}\rangle$ & Total \\
\hline
\multicolumn{5}{c}{{\it Ab initio} results}\\[0.3ex]
3  & 3  & $-1.1389$ & 3.3557 & 2.2168 \\
3  & 4  & ~~1.6080 & 6.0009 & 7.6089 \\
4  & 3  & ~~2.0464 & 4.7093 & 6.7557 \\
4  & 4  & $-1.2967$ & 3.8206 & 2.5239 \\[0.3ex]
\hline
\multicolumn{5}{c}{Semi-empirical results}\\[0.3ex]
3  & 3  & $-1.1391$ & 3.3376 & 2.1985 \\
3  & 4  & ~~1.6078 & 5.9681 & 7.5759 \\
4  & 3  & ~~2.0463 & 4.6834 & 6.7297 \\
4  & 4  & $-1.2969$ & 3.8000 & 2.5031 \\[0.3ex]  
\end{tabular}
\end{ruledtabular}    
\label{tab_main}
\end{table}

We also analyze contributions to the ${\cal X}_{PV}^{NSD}$ values arising from individual terms of the RCCSD method explicitly in Table \ref{tab3}. In this table, contributions arising through $\bar{D} T_1^{(1)}$ and its c.c. term correspond to the Core contributions and rest of the terms offer to the Valence contributions. From computational convenience, only the effective one-body part of $\bar{D}$ has been calculated and stored, which is multiplied with the RCC operators to get their values in the above table. The remaining part of $\bar{D}$ is coded directly along with the RCC operators and they are many in numbers. These contributions are referred to as the ``Others" in the above table. Again, wave functions of the RCC method are not normalized. We list corrections to the ${\cal X}_{PV}^{NSD}$ values due to normalization of the wave functions under `Norm' in the above table. In fact, these corrections are a part of the correlation effects in the system and should be taken into account for accurate estimate of the above quantities unless they have been absorbed in the formulation of the theory like normal RCC theory \cite{Sahoo2018PRL}. However, these corrections do not appear in other employed methods because in such cases either ket state or bra state is approximated to the DHF wave function and this is a limitation of those methods. Looking at the above table carefully, it can be observed that there are cancellations among the corrections from $\bar{D} T_1^{(1)}$ and its c.c. term in the $F_i=3 \rightarrow F_f=3$ and $F_i=4 \rightarrow F_f=4$ transitions, whereas they add up in the  $F_i=3 \rightarrow F_f=4$ and $F_i=4 \rightarrow F_f=3$ transitions. A similar trend is also seen in the contribution from the $\bar{D} S_{1i}^{(1)}$ and $S_{1f}^{(1)\dagger} \bar{D}$ terms. By drawing parallel to the calculation of the $E1_{PV}^{NSI}$ amplitudes, see Ref. \cite{Chakraborty2023}, it can be shown that $\bar{D} T_1^{(1)}$ and its c.c. term includes all Core contributions of the CPDF-RPA method in addition to contributions from PCs and many non-RPA type physical effects. It also includes DCP type Core correlation contributions which cannot appear through the CPDF-RPA method. Similarly, the Valence contributions of the CPDF-RPA* method are embedded within the $\bar{D} S_{1i}^{(1)}$ and $S_{1f}^{(1)\dagger} \bar{D}$ terms \cite{Chakraborty2023}. In addition, these RCC terms include valence PC and many non-RPA effects. As demonstrated explicitly in Ref. \cite{Chakraborty2023} in the analyses of the $E1_{PV}^{NSI}$ results, the DCP effects included in the CPDF-RPA method appear via the $\bar{D} S_{2i}^{(1)}$ and $S_{2f}^{(1)\dagger} \bar{D}$ RCC terms. It, therefore, implies that contributions from $S_{1f}^{(0)\dagger} \bar{D} S_{1i}^{(1)}$, $S_{1f}^{(1)\dagger} \overline{D} S_{1i}^{(0)}$ and other non-linear terms, quoted as `Others', are the additional correlation contributions arising through the RCCSD method.
\begin{figure*}[t]
\centering
\begin{tabular}{c c}\\
\includegraphics[width=70mm,height=60mm]{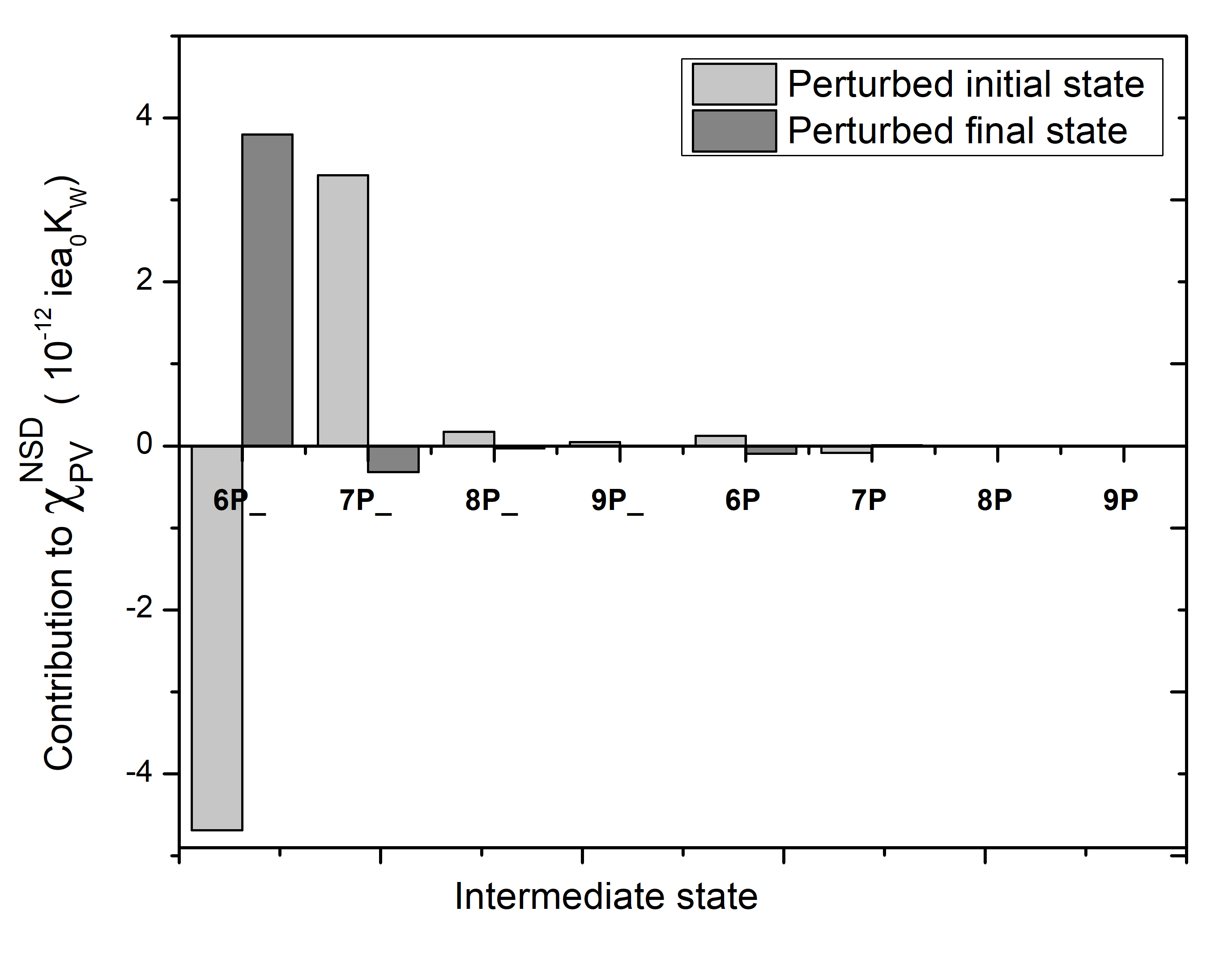} &
\includegraphics[width=70mm,height=60mm]{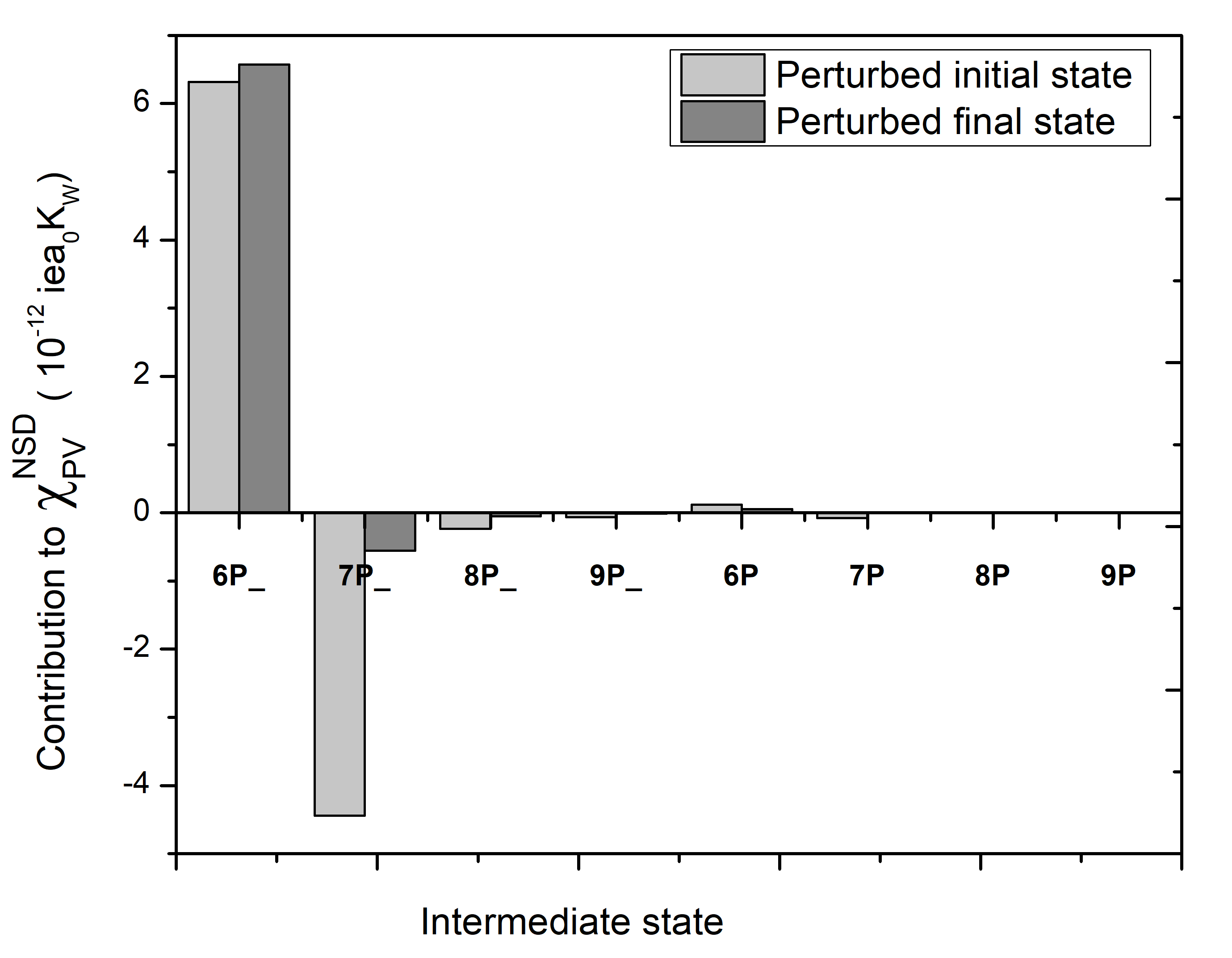}\\ 
(a) $F_f=3-F_i=3$ transition & (b) $F_f=3-F_i=4$ transition  \\
\includegraphics[width=70mm,height=60mm]{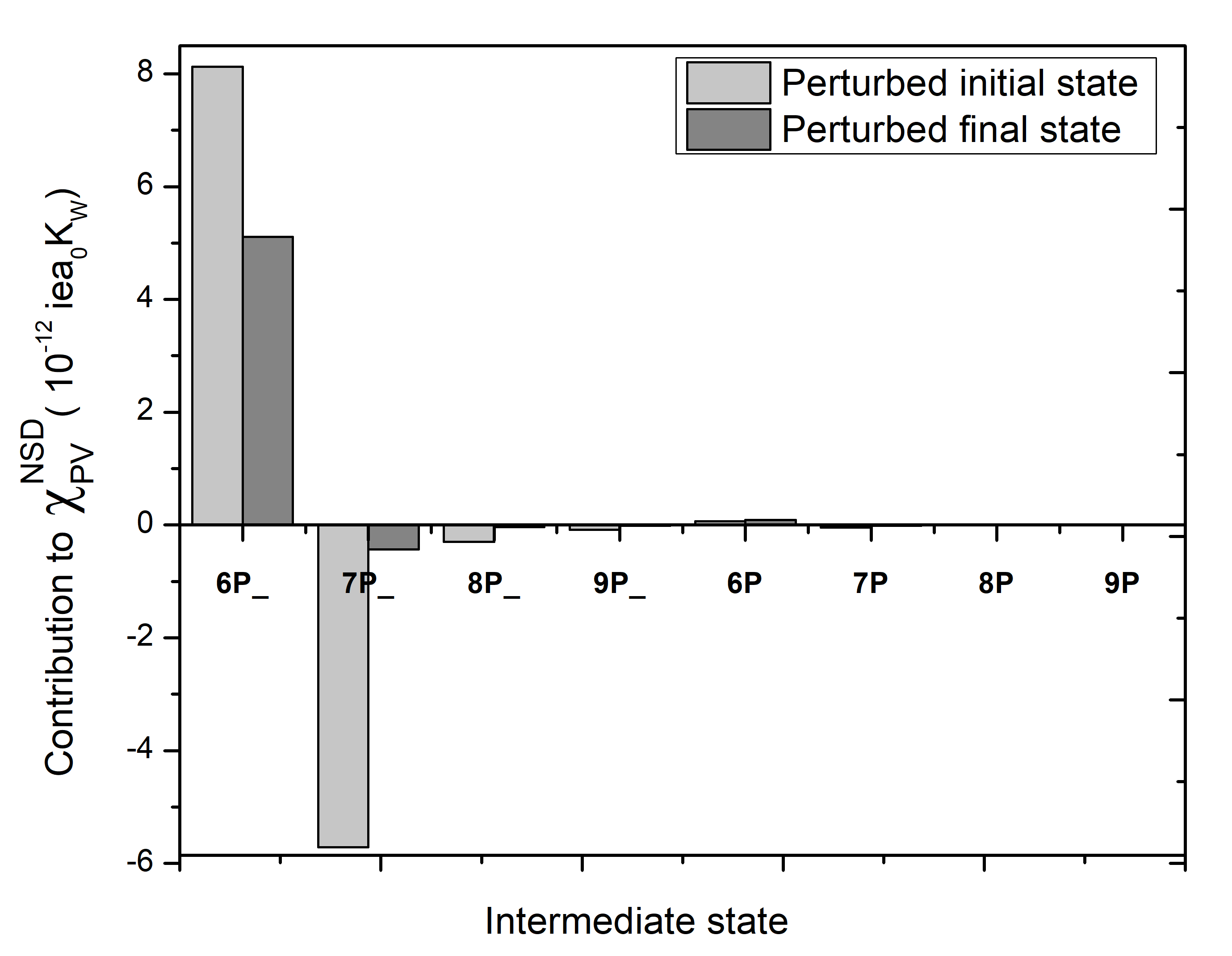} &
\includegraphics[width=70mm,height=60mm]{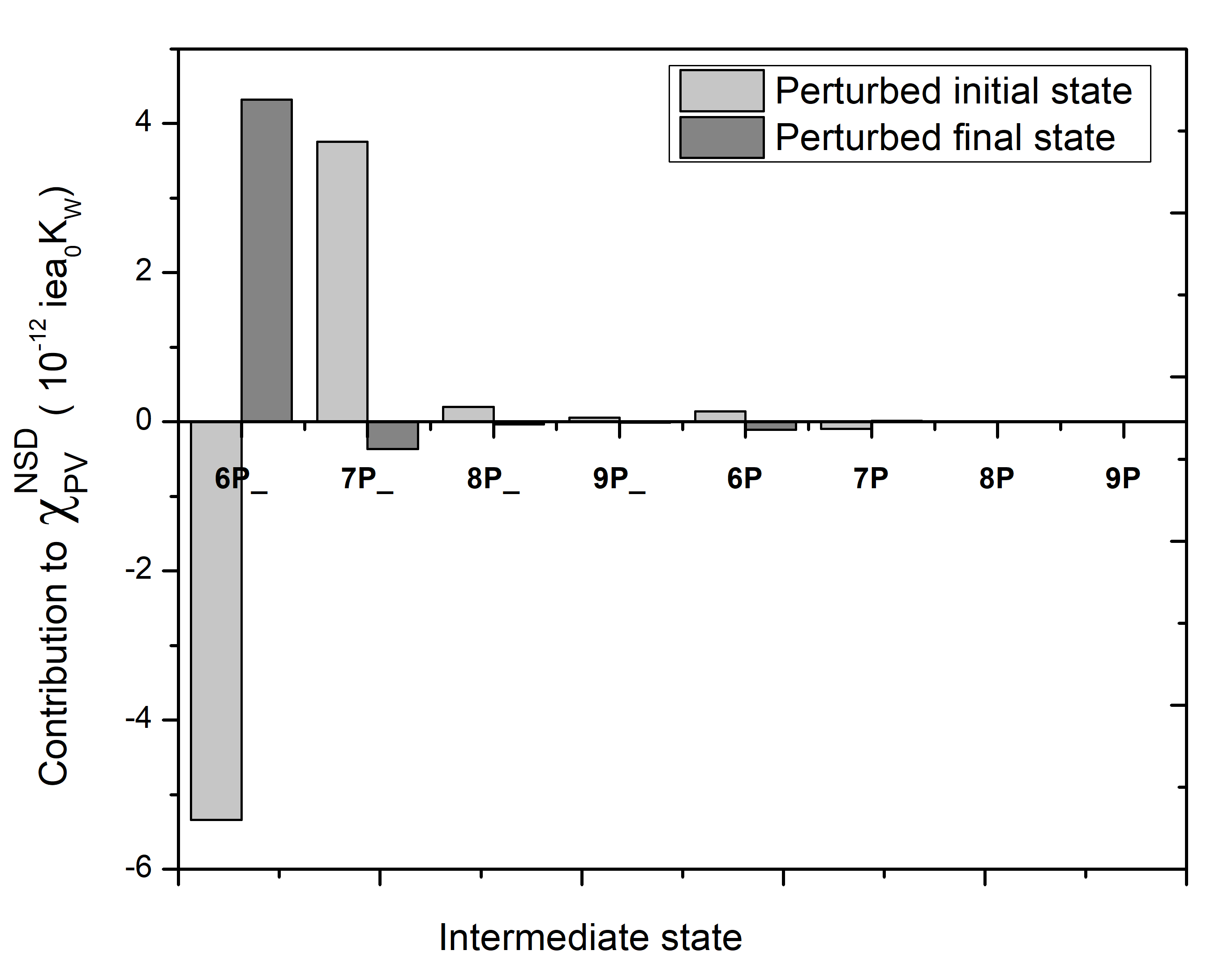}\\ 
(c) $F_f=4-F_i=3$ transition & (d) $F_f=4-F_i=4$ transition  \\
\end{tabular}
\caption{Demonstration of contributions from different intermediate states to the ${\cal X}_{PV}^{NSD}$ values for different $7s ~ ^2S_{1/2} (F_f) - 6s ~ ^2S_{1/2}(F_i)$ transitions in $^{133}$Cs. States with subscript `$-$' symbol in the figure represent the lower angular momentum state of a fine-structure partner; i.e. $P_{\_}$ means $P_{1/2}$ and $P$ stands for the $P_{3/2}$ state.}
\label{fig2}
\end{figure*}

As mentioned in Sec. \ref{secth}, all the above calculations were performed using the Fermi charge distribution. To get an impression of how these results vary with different nuclear charge distributions, we estimated the ${\cal X}_{PV}^{NSD}$ values using the uniform and Gaussian nuclear charge distributions \cite{Visscher1997, Andrae2000, Mitra2021} in the RCCSD method. We have presented these results in Table \ref{tab_nuc}. As can be seen from the table, there are differences about 0.5-0.6\% among the results from the Fermi and Gaussian charge distributions. However, these differences are comparatively large, about 1.6-1.8\%, among the results from the Fermi and uniform distributions. This suggests that the ${\cal X}_{PV}^{NSD}$ values strongly depend on the choice of the nuclear charge distributions. Since the Fermi charge distribution is more realistic than the Gaussian and uniform distributions, we have considered results from the Fermi charge distribution for the final values of ${\cal X}_{PV}^{NSD}$.

After learning trends of electron correlation effects to the evaluation of ${\cal X}_{PV}^{NSD}$ for all the transitions among the hyperfine levels of the $6s ~ ^2S_{1/2}$ and $7s ~ ^2S_{1/2}$ states in $^{133}$Cs from different angles using various methods and RCC terms, we finally wish to estimate and improve accuracy of these quantities so that they can be eventually used to infer fundamental parameters of general interest. This is done in two steps. In the first step, we try to separate out the Main contribution and the rest from our RCCSD calculation. This is because in the earlier results obtained using the sum-over-states approach, contributions from Core and Tail were estimated using lower-order methods. Moreover, DCP was not taken into account in those calculations. So by dividing RCCSD results into Main and the rest, the latter part will correspond to the Core, DCP, and Tail contributions together. Accuracy of this part can be easily claimed to be improved over the previous calculations. In the second step, it can be attempted to improve the accuracy of the Main contribution obtained through the first-principle approach using the RCCSD method. There is always a challenge to attain very accurate results in the {\it ab initio} method to calculate quantities like ${\cal X}_{PV}^{NSD}$ over the sum-over-states approach as the $K^{(1)}$ and E1 matrix elements converge very slowly with respect to higher-order correlation effects, while in the sum-over-states approach the important contributing E1 matrix elements can be used from the experiments. Also, one can use experimental energies in the sum-over-states approach to remove uncertainties due to the calculated energies. Since our ultimate objective of this work is to obtain very accurate values of ${\cal X}_{PV}^{NSD}$, we replace the {\it ab initio} Main contributions by the semi-empirical values using the sum-over-states approach. In the previous calculations using the sum-over-states approach \cite{Safronova2009}, the $np ~ ^2P_{1/2;3/2}$ (with $n=6-9$) intermediate bound states were used. To be consistent with these calculations and to demonstrate differences in the results from both the calculations, we also consider these intermediate states to estimate the Main contributions to ${\cal X}_{PV}^{NSD}$. First, we determine these quantities using the calculated E1 matrix elements and energies from the RCCSD method. This helps extract the Core, DCP, and Tail contributions from the RCCSD method. Then we replace the calculated E1 matrix elements and energies by the precisely known experimental values. In Table \ref{tab_elemnt}, we have given our calculated E1 matrix elements and energies from the RCCSD method, and compare them with the precisely reported experimental values \cite{Amiot2002, Damitz2019, Morton2000, Young1994, Toh2019, Toh2019-2}, while experimental energies are used from the National Institute of Science and Technology (NIST) database \cite{NIST}. For completeness, we have also listed the $K^{(1)}$ matrix elements from the RCCSD method in the same table. As can be seen from the table, there are significant differences between the RCCSD values and experimental results. In the previous study of $E1_{PV}^{NSI}$ in the $6s ~ ^2S_{1/2} \rightarrow 7s ~ ^2S_{1/2}$ transition of $^{133}$Cs \cite{Sahoo2021}, it has been shown that the RCCSD values of the E1 matrix elements and energies improve drastically towards the experimental results when triples excitations were taken into account in the RCC theory. Since triple excitations to determine the ${\cal X}_{PV}^{NSD}$ values are not implemented yet owing to very complex angular momentum couplings involved in this case, use of the sum-over-states approach to improve these results is the best possible option at this stage. It should be further noted that many E1 matrix elements have been reported precisely \cite{Damitz2019, Morton2000, Young1994, Toh2019, Toh2019-2} since the previous calculations of ${\cal X}_{PV}^{NSD}$ were carried out using the sum-over-states approach \cite{Safronova2009}. Therefore, the difference in the final results reported in Ref. \cite{Safronova2009} and the present values obtained using the sum-over-states approach can be partly due to this reason. We give contributions to the ${\cal X}_{PV}^{NSD}$ of different transitions involving the hyperfine levels of the $6s ~ ^2S_{1/2}$ and $7s ~ ^2S_{1/2}$ states from the $\it ab initio$ calculations in Table \ref{tab_main}. Again, these values are given by considering some of the E1 matrix elements and energies from the experiments (semi-empirical results) in the same table. We depict individual contributions from the $np ~ ^2P_{1/2,3/2}$ ($n=6-9$) states to the Main contribution in Fig. \ref{fig2}, which signifies the importance of their contributions. This knowledge can help improve the accuracy of the ${\cal X}_{PV}^{NSD}$ values in the future. The plots in the above figure show that among all the considered bound states, only the $6P_{1/2}$ and $7P_{1/2}$ states contribute almost entirely to the Main contributions in all transitions. As can be seen from the figure, contributions from the $nP_{3/2}$ states are very small. In the $F_i=3 \rightarrow F_f=3$ and $F_i=4 \rightarrow F_f=4$ transitions, the $6P_{1/2}$ state contributes in opposite sign to the initial and final perturbed states. As a result, the dominating contributions come effectively from the $7P_{1/2}$ state. In the $F_i=3 \rightarrow F_f=4$ and $F_i=4 \rightarrow F_f=3$ transitions, contributions from the $6P_{1/2}$ state are in the same sign for the both initial and final perturbed states. Therefore, contributions from the $6P_{1/2}$ state in these transitions dominate over the contributions from the $7P_{1/2}$ state. Also, there are huge cancellations among contributions from the intermediate states to both the initial and final perturbed states in all transitions. Particularly, these cancellations are strong in the initial perturbed state explaining the reason why contributions from the initial perturbed states are quite small over the final perturbed states.   

 It can be seen from Table \ref{tab_main} that the differences between the Main contributions from {\it ab initio} results and the final results given in Table \ref{tab0} are noticeable prominently. From this we can argue that accurate evaluation of the remaining contributions from Core, DCP, and Tail are equally important as the Main contributions. Again, the $K^{(1)}$ matrix elements are evaluated using the RCCSD method in this work. This attempt in the present work is to account for the Core, DCP, and Tail contributions to the ${\cal X}_{PV}^{NSD}$ values more accurately compared to the previous results reported using the sum-over-states approach. Then by adding the semi-empirical values of the Main contributions, we give the final recommended values of ${\cal X}_{PV}^{NSD}$ among the hyperfine levels of the $6s ~ ^2S_{1/2} \rightarrow 7s ~ ^2S_{1/2}$ transition in $^{133}$Cs, which can be combined further with the measured $E1_{PV}^{NSD}$ values to infer NAM. 

\begin{table}[t!]
\centering
\caption{The final recommended Main, Core, DCP$+$Tail contributions to the ${\cal X}_{PV}^{NSD}$ values (in units of $i e a_0 K_W \times10^{-12}$) of the $7s ~ ^2S_{1/2} (F_f) - 6s ~ ^2S_{1/2} (F_i)$ transitions in $^{133}$Cs. Uncertainties are quoted within the parentheses. Refer to the text for explanation on their estimations.}
\begin{ruledtabular}
\begin{tabular}{c c c c c c c}
$F_f$ & $F_i$ & Main & Core & DCP$+$Tail & Total\\
\hline
3 & 3 & 2.1985(52) & $-0.0047(2)$ & 0.122(3) & 2.316(6) \\  
3 & 4 & 7.5759(73) & $-0.3458(18)$ & 0.131(2) & 7.361(8) \\  
4 & 3 & 6.7297(88) & $-0.3441(18)$ & 0.084(1) & 6.470(9) \\  
4 & 4 & 2.5031(56) & $-0.0052(3)$ & 0.139(3) & 2.637(6)\\  
\end{tabular}
\end{ruledtabular}
\label{tab_rec}
\end{table}
 
In Table \ref{tab_rec}, we give the recommended values for the Main, Core, DCP, and Tail contributions to ${\cal X}_{PV}^{NSD}$, in units ie$a_0K_W\times10^{-12}$, along with their estimated uncertainties for all possible transitions among the hyperfine levels of the $6s ~ ^2S_{1/2}$ and $7s ~ ^2S_{1/2}$ states in $^{133}$Cs by adopting the procedure discussed in the previous two paragraphs. Since there is no obvious way of disengage the DCP and Tail contributions in our RCCSD method, they are given together as DCP$+$Tail in the above table. The major source of uncertainty to the Main contributions comes from the E1 matrix elements used in the experiments. We also repeated the calculations by increasing the size of basis functions in the CPDF-RPA method with different combinations of high-lying $s$ and $p$ orbitals. From the variations in the values from different correlation contributions, we assigned uncertainties to the Core and DCP$+$Tail contributions. Since carrying out such analyses using the RCCSD method is computationally cumbersome, we have used the CPDF-RPA method to estimate uncertainties to the Core and DCP$+$Tail contributions.  

Apart from $E1_{PV}^{NSD}$, there can be another NSD contribution to $E1_{PV}$ due to the hyperfine-induced NSI interaction if we consider next-order correction to $|(I J) F M_F \rangle$ by the hyperfine interaction Hamiltonian. We neglect this correction in the present work. It was mentioned earlier that the NSD component of $E1_{PV}$ between the hyperfine levels of the $6s ~ ^2S_{1/2} \rightarrow 7s ~ ^2S_{1/2}$ transition in $^{133}$Cs has been measured. However, it corresponds to a differential value of the NSD contribution to $E1_{PV}$ that was extracted by carrying out PV measurements between the $6s ~ ^2S_{1/2} (F_i=3) \rightarrow 7s ~ ^2S_{1/2} (F_f=4)$ transition and between the $6s ~ ^2S_{1/2} (F_i=4) \rightarrow 7s ~ ^2S_{1/2} (F_f=3)$ transition in $^{133}$Cs \cite{Wood1997}. By assuming the net NSD contribution to $E1_{PV}$ arises only from $E1_{PV}^{NSD}$ then we can express the differential $E1^{NSD}_{PV}$ value between the above hyperfine levels $F_i$ and $F_f$ as \cite{Johnson2003}
\begin{eqnarray}
    \delta E1^{NSD}_{PV}=K_W \left[ \left ( \frac{{\cal X}_{PV}^{NSD}}{{\cal A}_{F_f,F_i}} \right )^{F_f,F_i} - \left ( \frac{{\cal X}_{PV}^{NSD}}{{\cal A}_{F_i,F_f}} \right )^{F_i,F_f} \right] ,  \ \ \
\end{eqnarray}
where subscript and superscript $F_f,F_i$ notations used in the above expression denotes for the hyperfine level transition $F_f \rightarrow F_i$ and 
\begin{eqnarray}
{\cal A}_{F_fF_i} &=& (-1)^{J_f+F_i+I+1}\sqrt{6(2F_i+1)(2F_f+1)} \nonumber \\ && \times \begin{Bmatrix}
  F_f & F_i & 1 \\
  J_i & J_f & I
\end{Bmatrix}. \nonumber
\end{eqnarray}

Again, the actual measured quantity in Ref. \cite{Wood1997} is $\delta E1_{PV}^{NSD}/\beta=-0.077(11)$ $mV/cm$, where $\beta$ is the vector polarizability of the $6s ~ ^2S_{1/2} \rightarrow 7s ~ ^2S_{1/2}$ transition in $^{133}$Cs. Using the recently reported value $\beta=27.043(36)$ a.u. from Ref. \cite{Quirk2024}, we infer $K_W=0.116(16)$ by combining our calculated ${\cal X}_{PV}^{NSD}$ values with the measurement of $\delta E1_{PV}^{NSD}/\beta$. Furthermore by substituting $K_{NSD}=0.0140$ from a nuclear model calculation \cite{Haxton2001-2}, it yields $K_a=0.102(16)$. This is in good agreement with the values reported for $K_a$ in Refs. \cite{Johnson2003} and \cite{Haxton2001-2}. 

\section{Summary}

We have carried out calculations of the electric dipole amplitudes due to spin dependent parity-violating interactions in the $6s ~ ^2S_{1/2} \rightarrow 7s ~ ^2S_{1/2}$ transition of $^{133}$Cs by using the Dirac-Coulomb atomic Hamiltonian. To understand the roles of various correlation effects and contributions arising through many intermediate states to these quantities, we have employed the DHF, CPDF, RPA, and RCC methods. These contributions are further classified into Core, DCP, and Valence contributions, and their trends are analyzed using different methods. We have also analyzed contributions to the above quantities arising through various terms of the RCC method. Since the correlation effects arising through the other methods are implicitly present within the RCC method along with many other effects, we presume the results obtained using the RCC method are more accurate. To improve accuracy of these calculations further, we divide the Valence contributions of the RCC method into two parts: a part that comes from the low-lying bound states and the remaining part arises due to the other intermediate states. Among these two, the first part of the contributions is the dominant one. We improve the accuracy of this part by reevaluating them using a sum-over-states approach after replacing some of the precisely known electric dipole amplitudes and energies from experiments. Finally, we revise the limit on the magnitude of a nucleon-nucleon parity-violating coupling constant by combining our calculation with the precise measurement of the above transition in $^{133}$Cs.

\section*{Acknowledgment}

This work is supported by the Department of Space, Government of India. All the calculations reported in the present work were computed using the ParamVikram-1000 HPC cluster of the Physical Research Laboratory (PRL), Ahmedabad, Gujarat, India.

\end{document}